\documentclass[11pt]{article}
\pdfoutput=1
\usepackage{amsmath, amsfonts, amssymb}
\usepackage{comment}
\usepackage{graphicx}
\usepackage{psfrag}
\usepackage{amsthm}
\usepackage[usenames,dvipsnames,svgnames,table]{xcolor}
\usepackage{enumerate}
\usepackage{tcolorbox}
\usepackage{mathtools}
\usepackage{soul}
\usepackage{subfigure}
\usepackage{mathrsfs}  
\usepackage{a4wide}
\usepackage{booktabs}
\usepackage{tikz}
\usepackage{tikz-cd}
\usepackage{makecell}

\usepackage{color}
\definecolor{dark-gray}{gray}{0.20}
\definecolor{gray}{gray}{0.30}
\definecolor{light-gray}{gray}{0.80}
\definecolor{dark-red}{rgb}{0.7,0,0}
\definecolor{dark-green}{rgb}{0.1,0.4,0}
\definecolor{dark-blue}{rgb}{0.3,0.3,0.7}
\definecolor{light-blue}{rgb}{0.8,0.8,1}
\definecolor{swamp}{RGB}{240, 199, 197}

\usepackage{pifont}
\usepackage{setspace}

\newcommand{\be}{\begin{equation}}
\newcommand{\ee}{\end{equation}}

\def\be{\begin{equation}}
\def\ee{\end{equation}}
\def\bea{\begin{eqnarray}}
\def\eea{\end{eqnarray}}

\newcommand{\dd}{\mathrm{d}}

\def\simleq{\; \raise0.3ex\hbox{$<$\kern-0.75em
		\raise-1.1ex\hbox{$\sim$}}\; }
\def\simgeq{\; \raise0.3ex\hbox{$>$\kern-0.75em
		\raise-1.1ex\hbox{$\sim$}}\; }

\numberwithin{equation}{section}

\usepackage{jheppub} 
\hypersetup{
	colorlinks=true,
	linkcolor=dark-blue,
	citecolor=dark-red,
	urlcolor=dark-green,
	linktoc=page,
	pageanchor=false
}

\title{\centering
$\alpha'$ corrections to KPV:\\An uplifting story
}

\author{Simon Schreyer$^1$}
\author{ and Gerben Venken$^1$}

\affiliation{$^1$ Institute for Theoretical Physics, Heidelberg University,\\
	Philosophenweg 19, 69120 Heidelberg, Germany} 
\emailAdd{s.schreyer@thphys.uni-heidelberg.de}
\emailAdd{g.venken@thphys.uni-heidelberg.de}

\abstract{
In earlier work, the effect of $\alpha'^2$ curvature corrections on the NS5-brane responsible for the decay of anti-D3-branes in the set-up of Kachru, Pearson, and Verlinde (KPV) was considered. We extend this analysis to include all known $\alpha'^2$ corrections to the action of an abelian fivebrane which involve not just curvature but also gauge fields and flux. We compute the value of these terms at the tip of the Klebanov-Strassler throat to obtain the $\alpha'^2$ corrected potential for the NS5-brane of KPV. The resulting potential provides a novel uplifting mechanism where one can obtain metastable vacua with an arbitrarily small positive uplifting potential by fine-tuning $\alpha'$ corrections against the tree-level potential. This mechanism works for small warped throats, both in terms of size and contribution to the D3-tadpole, thereby sidestepping the issues associated with a standard deep warped throat uplift which are deadly in KKLT and, as we explicitly check, severely constraining in the Large Volume Scenario.
}

\begin{document}

\makeatletter
\let\old@fpheader\@fpheader

\makeatother

\maketitle


\section{Introduction}\label{intro}
The set-up of Kachru, Pearson, and Verlinde (KPV) \cite{Kachru:2002gs} provides one of the best-known examples of a proposed metastable spontaneously supersymmetry breaking vacuum in string theory. One considers $p$ anti-D3-branes at the tip of the Klebanov-Strassler (KS) throat \cite{Klebanov:2000hb}. These puff up into a single fluxed NS5-brane which depending on the parameters at the tip either has a metastable supersymmetry breaking vacuum or is classically unstable and pulled over the tip to decay into a supersymmetric vacuum. Our main interest in KPV is as a tool in the anti-D3-brane uplift to construct de Sitter vacua in string theory.

It remains one of the core questions of string phenomenology whether one can achieve controlled de Sitter vacua in string theory \cite{Danielsson:2018ztv,Obied:2018sgi}. One of the most well-trodden paths towards constructing de Sitter vacua is to start by constructing a scale separated AdS vacuum\footnote{This first step of constructing a landscape of scale-separated AdS vacua has also been called into question see e.g.~\cite{Sethi:2017phn,Gautason:2018gln,Lust:2019zwm,Lust:2022lfc,Cribiori:2022trc,Montero:2022ghl}. In the remainder of this paper we will proceed under the assumption that a landscape of scale-separated AdS vacua can be obtained in IIB string theory via either KKLT or LVS or some other mechanism, but this assumption is highly nontrivial.} in IIB with potential $V_\text{AdS}$. One then supplies a source of positive potential energy $V_\text{up}$ to provide an uplift to a de Sitter minimum. This is the route taken by e.g.~KKLT \cite{Kachru:2002gs} and LVS \cite{Balasubramanian:2005zx, Conlon:2005ki}. One requires that $|V_\text{up}| \approx |V_\text{AdS}|$, in order to obtain a metastable de Sitter vacuum. If $|V_\text{up}| \gg |V_\text{AdS}|$ the uplifting potential completely overpowers the AdS potential which provided a minimum and one obtains a runaway instead.

An anti-D3-brane, pointlike in the internal dimensions provides a source of positive potential energy but by itself would lead to a runaway. This is where the KS throat comes to the rescue: by placing the anti-D3-brane at the tip of a warped throat, one can exponentially suppress $V_\text{up}$ through warping. One may then set the warping at the tip of the throat such that $V_\text{up}$ provides a controlled uplift, as done in KKLT \cite{Kachru:2002gs} and LVS \cite{Balasubramanian:2005zx, Conlon:2005ki}. The issue with this is that in order to have sufficient warping one must have a sufficiently large throat, both in terms of its size and its contribution to the D3-tadpole. The issue of fitting such a large throat in the compactification geometry seems deadly in KKLT, where it has been dubbed the singular bulk problem \cite{Freivogel:2008wm,Carta:2019rhx, Gao:2020xqh}, and seems severely constraining in the LVS \cite{Junghans:2022exo, Gao:2022fdi, Junghans:2022kxg,Hebecker:2022zme}.

Recently, \cite{Hebecker:2022zme} shed new light on this issue by examining how $\alpha'$ corrections affect uplifts with warped throats. Such corrections become important when one considers small warped throats which should be easier to embed in a compactification geometry. In particular, the effect of curvature corrections on the worldvolume of the uplifting antibrane was examined, focussing especially on the impact of these curvature corrections on the KPV decay channel. This analysis opened up many interesting new possibilities. Most exciting was the possibility of a new uplifting mechanism unique to warped throats with a small tip. Curvature corrections correct the potential for a single antibrane as \cite{Bachas:1999um,Junghans:2014zla,Junghans:2022exo}
\begin{equation}
   V_{\overline{D3}} = \mu_3 \text{e}^{-\phi}\left[ 1 - \frac{(4 \pi^2 \alpha')^2}{384 \pi^2} R_{a \alpha b}^{\quad \alpha} R_{\;\beta}^{a \;\; b \beta}\right] = \mu_3 \text{e}^{-\phi}\left[ 1 - \frac{c}{(g_s M)^2}\right]\,,
\end{equation}
with $c=5.924$ and where $a$, $b$ are normal and $\alpha$, $\beta$ tangent indices relative to the brane. Here $M$ is the $F_3$ flux at the tip of the throat and the radius $R_{S^3}$ of the tip is proportional to $\sqrt{g_sM}$. By fine-tuning $g_s$ one can achieve $c/(g_s M)^2 \approx 1$ and in this way tune $V_{\overline{D3}}$ to an arbitrarily small value. In this way one can achieve an uplift without requiring a large warped throat. We refer to \cite{Hebecker:2022zme} for further discussion.

It is insufficient to analyze this uplift purely in terms of the anti-D3-brane. The antibrane puffs up and to study the metastable vacuum we must consider the KPV set-up. Unfortunately, the analysis of \cite{Hebecker:2022zme} had to remain very incomplete on this point as it only considered curvature corrections. The goal of the present paper is twofold: First, to take into account corrections to the worldvolume theory due to background fluxes which were neglected in \cite{Hebecker:2022zme}. Second, to more fully explore the phenomenological impact of $\alpha'$ corrections on the brane worldvolume in the KPV set-up.

The rest of the paper is structured as follows. In Sect.~\ref{sec:review} we very briefly review the set-up of KPV \cite{Kachru:2002gs} and how the results of \cite{Hebecker:2022zme} affect this set-up.

In Sect.~\ref{sec:fluxcorretions} we consider additional corrections from flux, gauge fields and mixed terms involving both of the above as well as curvature. We compute the potential for the NS5-brane with these corrections taken into account. We show that for throats with a large tip the analysis of KPV is recovered, while for small throats the corrections become important and a sensible potential with interesting new features appears.

In Sect.~\ref{sec:newuplift} we show that using these new features one can indeed get the new uplifting mechanism by finetuning $\alpha'$ corrections to work, while this was previously only hinted at. As this uplifting mechanism requires balancing $\alpha'$ corrections to the potential to be as large as the tree-level potential one is at the borderline of control. This novel uplifting mechanism allows one to uplift using a small warped throat, requiring a warped throat which contributes as little as $N=40$ to the D3-tadpole.

In Sect.~\ref{sec:deepthroat} we instead consider the traditional uplifting mechanism where one uses a large warped throat to exponentially warp down $V_{\text{up}}$. We investigate under what conditions this traditional uplift is controlled for the LVS specifically. Even with the weakest demands on control one finds that the warped throat must have a D3-tadpole $N=560$. Demanding a bit more control, one rapidly finds that an $\mathcal{O}(10^3)$ tadpole from the warped throat. This severely restricts the possibility of achieving such an uplift with a large warped throat as compactification geometries with sufficient negative D3-tadpole to fit such throats are currently not known in IIB string theory with local D7-tadpole cancellation (but larger tadpoles may be possible for nonlocal cancellation, see \cite{Crino:2022zjk})\footnote{Note that non-local D7-tadpole cancellation leads to additional corrections. See \cite{Junghans:2022exo,Junghans:2022kxg} for the conservative view concerning how these corrections affect the LVS.}.

Our paper focuses on $\alpha'$ corrections to the worldvolume action of an NS5-brane. However, many of the corrections we consider have strictly only been derived for a D5-brane and we have inferred the analogous correction for an NS5-brane. While we believe our method for inferring these corrections is correct, one may object that we strictly lack a derivation of the $\alpha'$ corrections on the NS5-brane (see the nontrivial working assumptions in App.~\ref{sec:SdualAppendix}). The main reason why we have focused on the KS throat is that this is the set-up most commonly considered for a warped throat uplift in the literature. One may instead consider the S-dual throat with a D5-brane at the tip and use this for the uplift \cite{Gautason:2016cyp}. One can then perform our analysis of $\alpha'$ corrections for this dual set-up now being confident that one knows the $\alpha'$ corrections. We have analyzed our corrections to this S-dual set-up in Sect.~\ref{sec:sdks}. Unfortunately the novel uplifting mechanism we propose does not work in this S-dual set-up with the main issue being that the radius of the tip of the throat is set by $\sqrt{K}$, with $K$ the amount of $H_3$ flux at the tip of the S-dual throat. We then lack a  factor of $\sqrt{g_s}$ which allows us to tune the radius of the tip.

We conclude in Sect.~\ref{sec:conclusions}. The new uplifting mechanism we propose requires far smaller throats than the traditional uplifting mechanism via exponential warping. This allows us to sidestep the issues involved in uplifting KKLT and LVS with a large warped throat. Some questions remain on how controlled our new uplifting mechanism is and we discuss future directions to gain a better understanding of the small throat uplifting mechanism.

The paper contains three appendices. In App.~\ref{sec:fluxcorrections} we gather various technical computations related to $\alpha'$ corrections. In App.~\ref{sec:TadpoleAppendix} we review the Parametric Tadpole Constraint (PTC) \cite{Gao:2022fdi} and recast it into a form more useful for applying the bounds on control for the warped throat we obtain to the LVS. We also comment on the discrepancy in the conditions for control in the LVS between \cite{Junghans:2022exo,Junghans:2022kxg} and \cite{Gao:2022fdi}.
In App.~\ref{sec:SdualAppendix} we discuss in detail how we infer the $\alpha'^2$ corrections on the NS5-brane from the corresponding corrections on D-branes while stressing the nontrivial assumptions underlying our argument.

\section{Review of KPV and curvature corrections}
\label{sec:review}

In the set-up of KPV \cite{Kachru:2002gs}, $p$ anti-D3-branes are placed at the tip of the Klebanov-Strassler (KS) throat \cite{Klebanov:2000hb}. The KS throat is a deformed conifold with $M$ units of $F_3$ flux threading the A-cycle of the throat. Topologically, at a distance $\tau$ from the tip there is a $T^{1,1}$ geometry, which is an $S^2$ fibration over an $S^3$. At the tip $\tau=0$ the $S^2$ shrinks to zero size while the $S^3$ remains of finite radius $R_{S^3}\sim \sqrt{g_sM}$ and coincides with the A-cycle.

\begin{figure}
    \centering
    \includegraphics[width=0.8\textwidth]{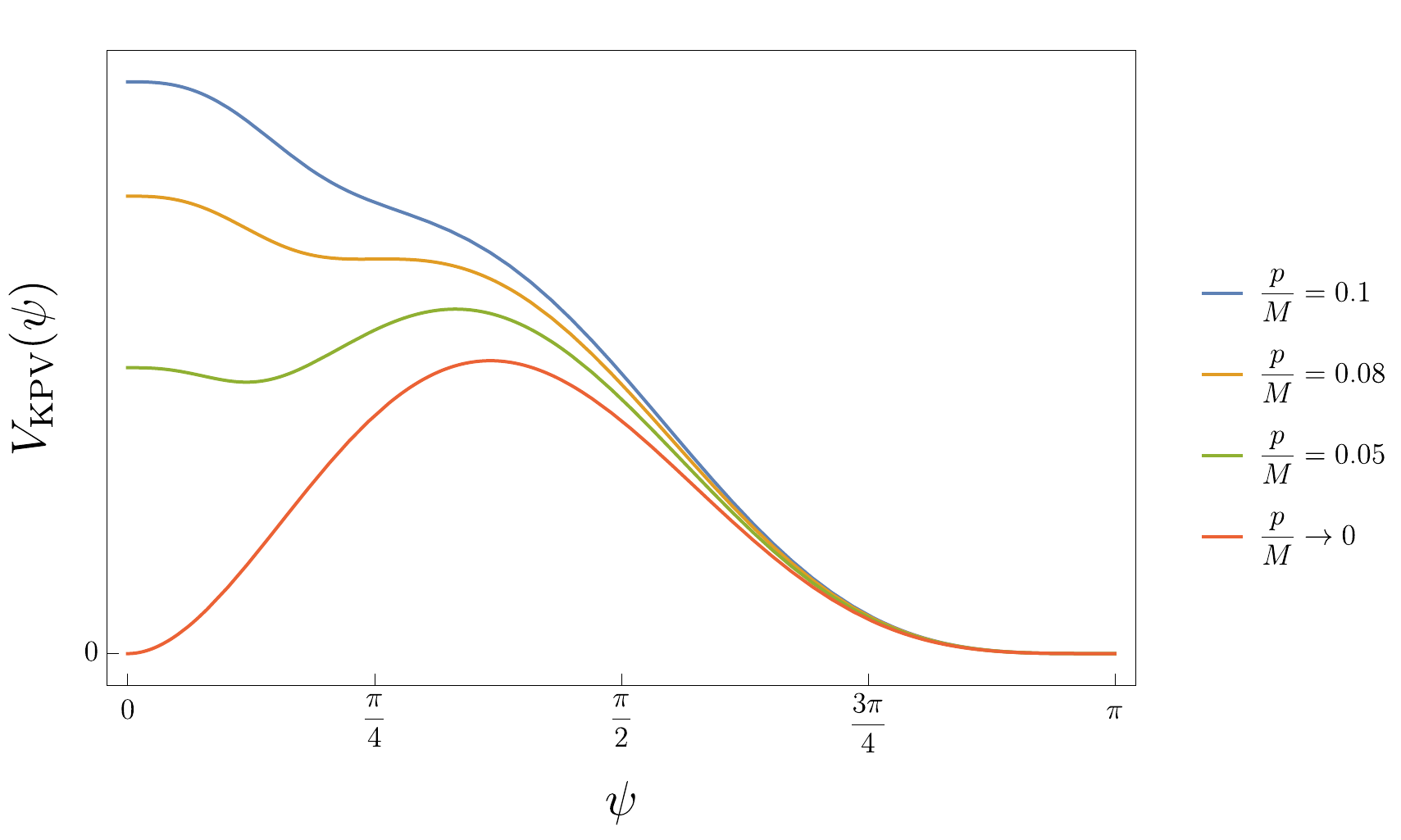}
    \caption{The potential $V_{\text{KPV}}(\psi)$ (suitably normalized) for different values of $p/M$.}
    \label{fig:KPVpot}
\end{figure}

The anti-D3-branes at the tip can puff up into an NS5-brane. The NS5 wraps an $S^2$ inside the $S^3$. The authors of \cite{Kachru:2002gs} calculated the 4d potential of this NS5-brane in terms of $M$, $p$, and $\psi$, the angular direction of the $S^3$ normal to the $S^2$ wrapped by the NS5, from the tree-level string-frame NS5-brane action
\begin{equation}
    S =  -\frac{\mu_{5}}{g_s^2} \int_{\mathcal{M}_{6}} \dd^{6}\xi \,  \sqrt{-\det\left(g_{\mu\nu}+  2\pi\alpha' g_s F_{2\, \mu \nu} - g_s C_{2 \, \mu \nu}\right)} - \mu_5 \int_{\mathcal{M}_{6}} B_6\,,
    \label{eq:ns5action}
\end{equation}
where $\mu_5$ is the brane tension.
The resulting potential is of the form
\begin{equation}
    \hspace*{-.2cm}V_{\text{KPV}}(\psi) = \frac{4 \pi^2 p \mu_5}{g_s} + \frac{4 \pi \mu_5 M}{g_s}\left(\sqrt{b_0^4 \sin^4 (\psi) + \left( \frac{\pi p}{M} -\psi +\frac{\sin(2\psi)}{2} \right)^2} -\psi +\frac{\sin(2\psi)}{2} \right)\hspace*{-.2cm}\,,
    \label{eq:vkpv}
\end{equation}
where $b_0^2\approx 0.93266$. The potential describes a decay channel for the NS5 into $M-p$ D3-branes. Whether the NS5 is metastable or unstable in this decay channel depends on the value of $p/M$. During this process, the NS5 slips over the equator while the anti-D3-brane charge of the NS5 annihilates against 3-form flux. The potential is shown in Fig.~\ref{fig:KPVpot} where one can see that for $p/M>0.08$ no metastable minimum exists. 

The main idea of \cite{Hebecker:2022zme} was to include curvature corrections to the potential of the NS5 since in phenomenologically viable models, the value $g_sM^2$ and hence $R_{S^3}\sim \sqrt{g_s M}$ is typically not too large. Thus, curvature corrections, scaling like $1/R_{S^3}^4$, eventually become important. This is a manifestation of the fact that the validity of the supergravity expansion in $\alpha'$ is on the borderline of control. The curvature corrections evaluated for the NS5 at the tip of the throat lead to a correction of the KPV potential
\begin{equation}
\begin{split}
    V_{\text{curv}}(\psi) =
    & -\frac{ c_1 +  c_2 \cot^2\psi(2+\cot^2\psi)}{(g_s M)^2}\frac{4 \pi \mu_5 M}{g_s} \sqrt{b_0^4 \sin^4 (\psi) + \left(\frac{\pi p}{M} -\psi +\frac{\sin(2\psi)}{2} \right)^2}\,,
    \end{split}
    \label{eq:vcurv}
\end{equation}
where $c_1\approx 8.825$ and $c_2\approx 1.891$. The terms $\sim \cot\psi$ come from the second fundamental form $\Omega$ that is non-zero for non-geodesically embedded branes. 

As was repeatedly stressed in \cite{Hebecker:2022zme}, great care had to be taken in interpreting this corrected potential and its regime of validity. One of the main issues is that only pure curvature $\alpha'^2$ corrections to the NS5 worldvolume action were included in this potential while at the same order in $\alpha'$ there are also corrections depending on gauge fields, field strengths, and mixing of these with curvature\footnote{We will generically refer to such corrections as flux corrections, even though they involve not just $F_3$, $H_3$, $F_5$ and $H_7$ but also $C_2$ and $F_2$.}. The main goal of this paper is to account for these additional corrections and we shall see that the resulting potential is much more readily physically interpretable.

\section{Flux corrections}

\subsection{Flux corrections on the NS5-brane}
\label{sec:fluxcorretions}

As described in Sect.~\ref{sec:review}, in \cite{Hebecker:2022zme} higher order $\alpha'$ corrections to the Chern-Simons action and corrections to the DBI action that involve fluxes were neglected. 
In this section, we summarize all flux corrections to D-branes in superstring theory from the papers \cite{Garousi:2009dj,Garousi:2010ki,Garousi:2010rn,Garousi:2011ut,Robbins:2014ara,Garousi:2014oya,Jalali:2015xca,Garousi:2015mdg,Jalali:2016xtv,BabaeiVelni:2016srs,Garousi:2022rcv} that are non-vanishing for the KS throat, extend them to the NS5-brane and evaluate them at the tip of the KS throat. Many of these corrections have only been derived in flat space without background $F_2$. We make the following assumptions to extend these corrections to a brane worldvolume with nonzero $F_2$ in a curved background. We promote all partial derivatives to covariant ones in order for the action to be Lorentz covariant. Moreover, we assume all corrections to hold also for non-zero $F_2$ and $B_2$ flux by replacing $\sqrt{-g}\to\sqrt{-(g+2\pi\alpha'\mathcal{F}_2)}$.

Schematically, the corrections to the DBI action of a D$p$-brane important for the KS throat read \cite{Robbins:2014ara,Garousi:2014oya,BabaeiVelni:2016srs,Garousi:2022rcv}
\begin{equation}
\begin{split}
    S_{\text{DBI,D}p} \supset \frac{\mu_p\alpha'^2}{g_s} & \int_{\mathcal{M}_{p+1}}\dd^{p+1}x \sqrt{-(g+2\pi \alpha' \mathcal{F}_2)} \Biggl[ H_3^4 + H_3^2 R \\
    &+ \Omega^4(2\pi\alpha'\mathcal{F}_2)^2 + (2\pi\alpha'\mathcal{F}_2) \Omega^2 \nabla H_3\Biggr]\,,
    \label{eq:dbidp}
    \end{split}
\end{equation}
where we omitted any index contraction and numerical prefactor, and $R$ symbolically represents either the Riemann or the Ricci tensor. The precise action can be found in App.~\ref{sec:nonzeroflux}, in \eqref{oplaneterms}, \eqref{eq:omega4f2} and \eqref{eq:dbif2omegah3}. Note that the first two terms in \eqref{eq:dbidp} are derived for O$p$-planes. At order $\alpha'^2$, up to overall factors, couplings on O-planes can be obtained from couplings on D-branes by orientifold projection \cite{Mashhadi:2020mzf}\footnote{Note that this is not true for higher order $\alpha'$ corrections, as can be seen from the expansion of anomalous CS couplings.}. From this we conclude that all couplings on O-planes at order $\alpha'^2$ are also present on D-branes.

We want to calculate the effect of $\alpha'$ corrections to the KPV process. Therefore we need to translate the $\alpha'$ corrections on D5-branes to NS5-branes. Here, we only state the results but the detailed argument on how to accomplish this is given in App.~\ref{sec:SdualAppendix}. Note that this argument is not a complete proof but relies on several non-trivial assumptions discussed in detail in App.~\ref{sec:SdualAppendix}. Crucially, one must assume an order-by-order in $\alpha'$ matching between the action of the stack of nonabelian D3-branes being puffed up and the action of the nonabelian fivebrane into which the stack puffs up. This matching is not guaranteed as the regimes where the threebrane perspective or the fivebrane perspective are controlled are different. As we discuss in Sect.~\ref{sec:sdks}, one could also work in the S-dualized KS throat \cite{Gautason:2016cyp} where the anti-branes puff up into a fluxed D5-brane and the $\alpha'^2$ corrections are known. 

To leading order in $g_s$, the terms on the NS5-brane corresponding to the terms \eqref{eq:dbidp} on the D5-brane are of the form
\begin{equation}
\begin{split}
    S_{\text{DBI,NS}5} \supset \frac{\mu_5}{g_s^2}\alpha'^2 &\int_{\mathcal{M}_{6}}\dd^6x \sqrt{-(g+2\pi\alpha' g_s\mathcal{F}_2)} 
    \Biggl[ (-g_sF_3)^4 + (-g_sF_3)^2 R \\
    &+ \Omega^4(2\pi\alpha'g_s\mathcal{F}_2)^2 + (2\pi\alpha'g_s\mathcal{F}_2)\Omega^2 \nabla(-g_sF_3) \Biggr]\,,
    \label{eq:dbins5}
    \end{split}
\end{equation}
where $2\pi\alpha' g_s \mathcal{F}_2 = 2\pi\alpha'g_s F_2 -g_s C_2$.
The overall $g_s$ prefactor in \eqref{eq:dbins5} matches the $1/g_s^2$ prefactor known from the tree level NS5-brane action \eqref{eq:ns5action} as expected. Note that in order to properly count the $g_s$ scaling it is important that the kinetic terms of all fields have the same $g_s$ dependence. We normalize all flux kinetic terms in such a way that they have a $g_s^{-2}$ dependence. 
This means for instance for $F_3$ that we write the kinetic term $\int_{10} |F_3|^2\sim g_s^{-2}\int_{10} |F_{3,\text{norm}}|^2$ with $F_{3,\text{norm}}= g_s F_3$. Similarly, we find $\mathcal{F}_{2,\text{norm}}= g_s \mathcal{F}_2$. This explains why $F_3$ and $\mathcal{F}_2$ always appear with a factor of $g_s$ in \eqref{eq:dbins5}.

Besides $\alpha'$ corrections to the DBI action, there are also $\alpha'$ corrections to the CS action. The important couplings for the KS throat schematically read \cite{Jalali:2016xtv}
\begin{equation}
    S_{\text{CS,D}5} \supset \mu_5 \alpha'^2 \int_{\mathcal{M}_{6}}\dd^{6}x\left(- \epsilon_{(6)} \mathcal{F}_2 R \nabla \tilde{F}_5 +\epsilon_{(6)}\mathcal{F}_2\nabla H_3\nabla F_7 \right)\,,
    \label{eq:csdp}
\end{equation}
where again we have omitted any index structure. Note that $\epsilon_{(6)}$ denotes the 6d Levi-Civita-symbol, $\tilde{F}_5$ the self dual 5-form flux and $F_7$ the 7-form flux. The numerical prefactors and index structures are given by \eqref{eq:csf5} and \eqref{eq:csh7} in App.~\ref{sec:nonzeroflux}. 
The corresponding action on the NS5-brane to leading order in $g_s$ is given by (see again App.~\ref{sec:SdualAppendix} for details)
\begin{equation}
    S_{\text{CS,NS}5} \supset \frac{\mu_5}{g_s^2} \alpha'^2 \int_{\mathcal{M}_{{6}}}\dd^{{6}}x\left(- \epsilon_{(6)} (g_s\mathcal{F}_2) R \nabla (g_s\tilde{F}_5) +\epsilon_{(6)}(g_s\mathcal{F}_2)\nabla (-g_sF_3)\nabla(g_s^2 H_7) \right)\,,
    \label{eq:csns5}
\end{equation}
where we have already properly accounted for factors of $g_s$ to obtain normalized kinetic terms. Let us compare the $g_s$ scaling in \eqref{eq:csns5} with the tree level term $\int\dd^6x \mathcal{F}_2\wedge C_4 = g_s^{-2}\int\dd^6x (g_s\mathcal{F}_2)\wedge (g_s C_4)$ present on the NS5-brane. This shows that the action \eqref{eq:csns5} appears again at the same order in $g_s$ as the tree level action. 

We have accounted for all $\alpha'^2$ corrections to the fivebrane worldvolume action we found in the literature. However, there is to our knowledge no proof that the $\alpha'^2$ corrections are completely known. Especially corrections to the DBI action of D$p$-branes involving $F_{(p)}$ flux are not known.
Until the fivebrane action at order $\alpha'^2$ is completely determined, our results must necessarily remain incomplete.

\subsection{The flux corrected KPV potential}

With the results of App.~\ref{sec:nonzeroflux}, the $\alpha'$ corrected KPV potential of all currently known corrections is given by \eqref{eq:vkpv}, \eqref{eq:vcurv}, and \eqref{eq:vflux}. It is of the form
\begin{equation}
\begin{split}
\label{eq:vtot}
    V_\text{tot} = &\, V_\text{KPV} + V_\text{curv} + V_\text{flux}\\
    = & \,\frac{4 \pi \mu_5 M}{g_s} 
      \sqrt{b_0^4 \sin^4 (\psi) + \left(p \frac{\pi}{M} -\psi +\frac{1}{2}\sin(2\psi) \right)^2}
       \times\Biggl[1+\frac{1}{(g_sM)^2} \Biggl( c_3 - c_1 \\
       & \qquad+ (c_4-2c_2) \cot^2\psi-c_2\cot^4\psi+\frac{ c_5 \cot^4\psi  }{ \sin^4\psi}\left(\frac{\pi p}{M}  -\left(\psi-\frac{\sin(2\psi)}{2}\right)\right)^2 \\
       & \qquad -\frac{c_6\,\cot^3\psi}{ \sin^2\psi}\left(\frac{\pi p}{M}  -\left(\psi-\frac{\sin(2\psi)}{2}\right)\right) \Biggr)\Biggr]\\
       & \qquad+\left[\frac{4\pi^2 p \mu_5}{g_s} -\frac{4 \pi \mu_5 M}{g_s} \left(\psi-\frac{\sin(2\psi)}{2}\right) \right] \left( 1+ \frac{c_7}{(g_sM)^2} + \frac{c_8 \cot\psi}{(g_sM)^2 \sin\psi}  \right)\,,
     \end{split}
\end{equation}
where the numerical constants $c_1$ to $c_8$ are given in Tab.~\ref{tab:cs}.

\begin{table}[!h]\centering
	\caption{Numerical constants in the $\alpha'$ corrected KPV potential $V_\text{tot}$.}
	\vspace{.3cm}
	\label{tab:cs}
		\begin{tabular}{cccccccc}
		\toprule
		$c_1$ & $c_2$ & $c_3$ & $c_4$ & $c_5$ & $c_6$ & $c_7$ & $c_8$ \\
			\midrule
			8.825&1.891 &2.448 &8.696 &32.61 &2.174 & 14.64& 18.65\\
			\bottomrule
	\end{tabular}
\end{table}

It was already observed in \cite{Hebecker:2022zme} that the curvature corrected KPV potential appears to have two parameters $g_sM$ and $p/M$ that determine the shape of the potential. Non-trivially, this is still true when the flux corrections are included:
All terms precisely line up in terms of $g_sM$ and $p/M$ with an overall $1/g_s^2$ factor.
Hence, the $\alpha'$ expansion of the brane action translates into an expansion of the potential in terms of $g_sM$ and $p/M$. The expansion in $p/M$ is a new effect not observed in \cite{Hebecker:2022zme}. It stems from corrections including $F_2$. 
The $F_2$ corrections in \eqref{eq:dbins5} are higher order in $\alpha'$ compared to the other corrections. Nevertheless, it still makes sense to include them into the potential since they feature the same $g_sM$ suppression but are additionally suppressed by the second expansion parameter $p/M$.

\begin{figure}
    \centering
    \includegraphics[width=\textwidth]{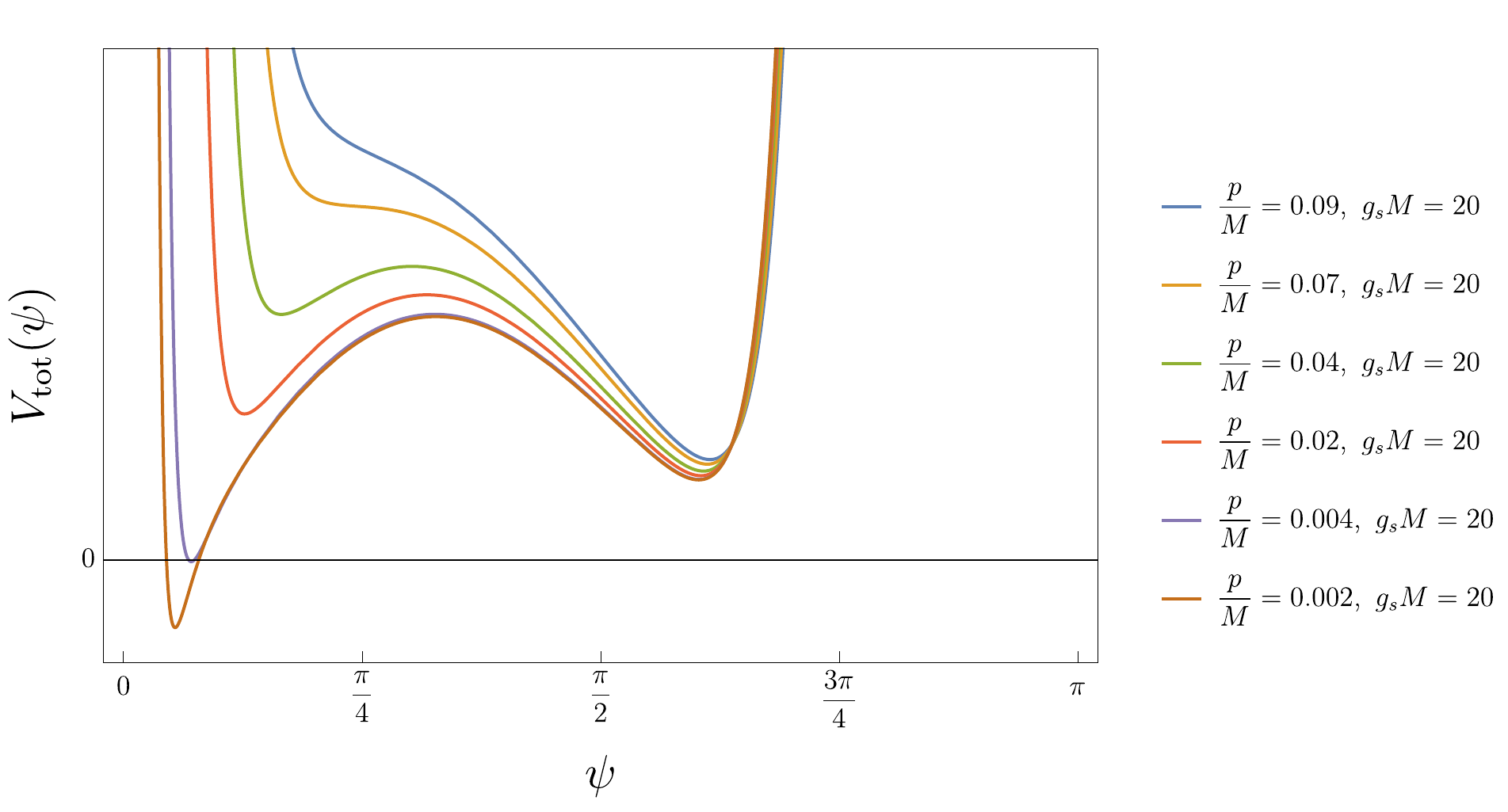}
    \caption{The potential $V_{\text{tot}}(\psi)$ (suitably normalized) for $g_sM=20$ and different values of $p/M$. With decreasing values of $p/M$, the form of the potential transitions from no metastable minimum to a minimum with $V_\text{up}<0$. In between, the vacuum energy is zero or positive. Note that we lose control over the perturbation theory that leads to this potential when either $\psi \gtrsim \pi/2$ or $\sqrt{g_s M}\sin \psi \lesssim 1$.}
    \label{fig:totpotvarypoverM}
\end{figure}

\begin{figure}
    \centering
    \includegraphics[width=\textwidth]{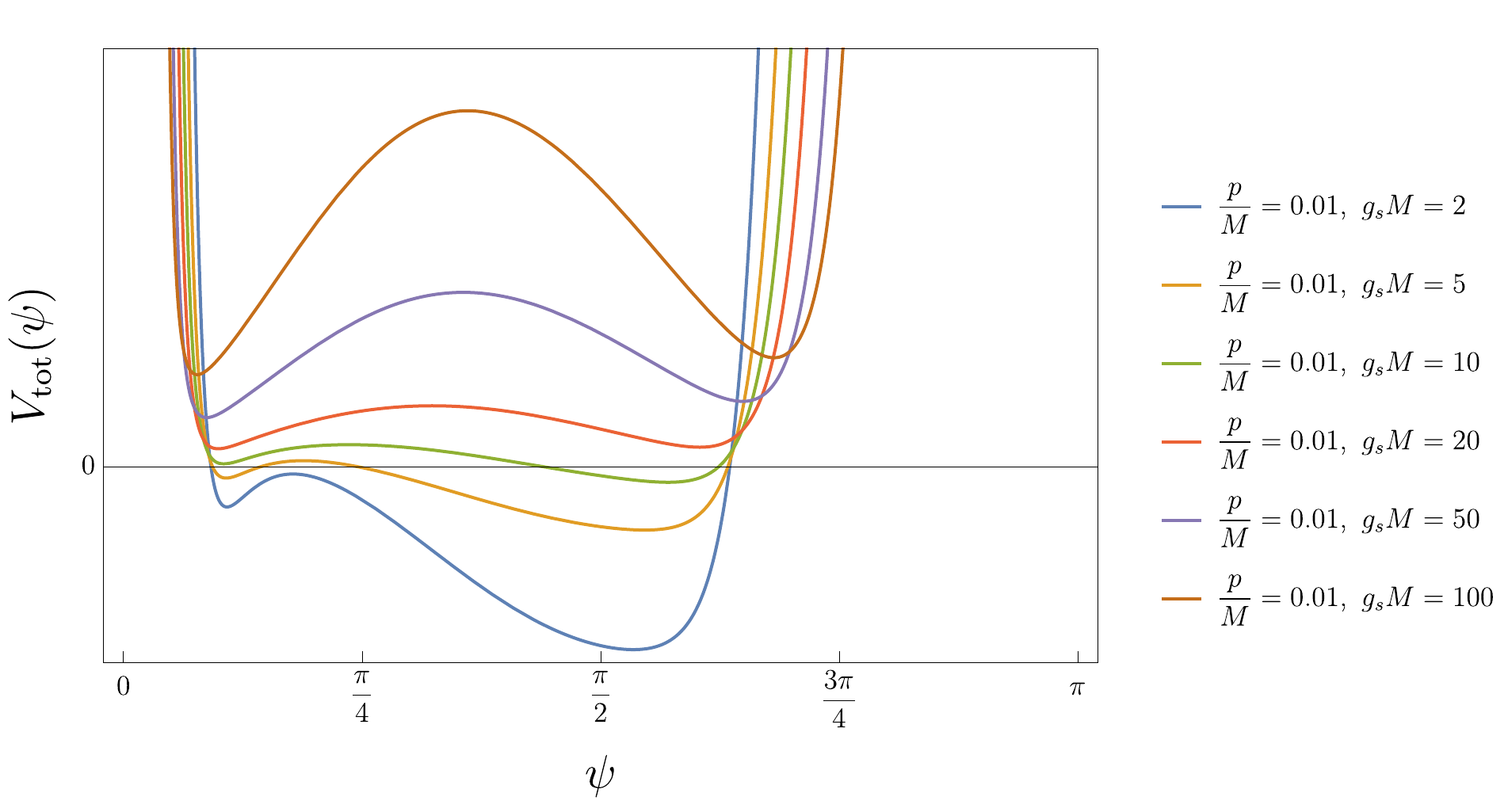}
    \caption{The potential $V_{\text{tot}}(\psi)$ (suitably normalized) for $p/M=0.01$ and different values of $g_sM$. Note that we lose control over the perturbation theory that leads to this potential when either $\psi \gtrsim \pi/2$ or $\sqrt{g_s M}\sin \psi \lesssim 1$.}
    \label{fig:totpotvarygsM}
\end{figure}

\begin{figure}[h]
    \centering
    \subfigure[]{\includegraphics[width=0.46\textwidth]{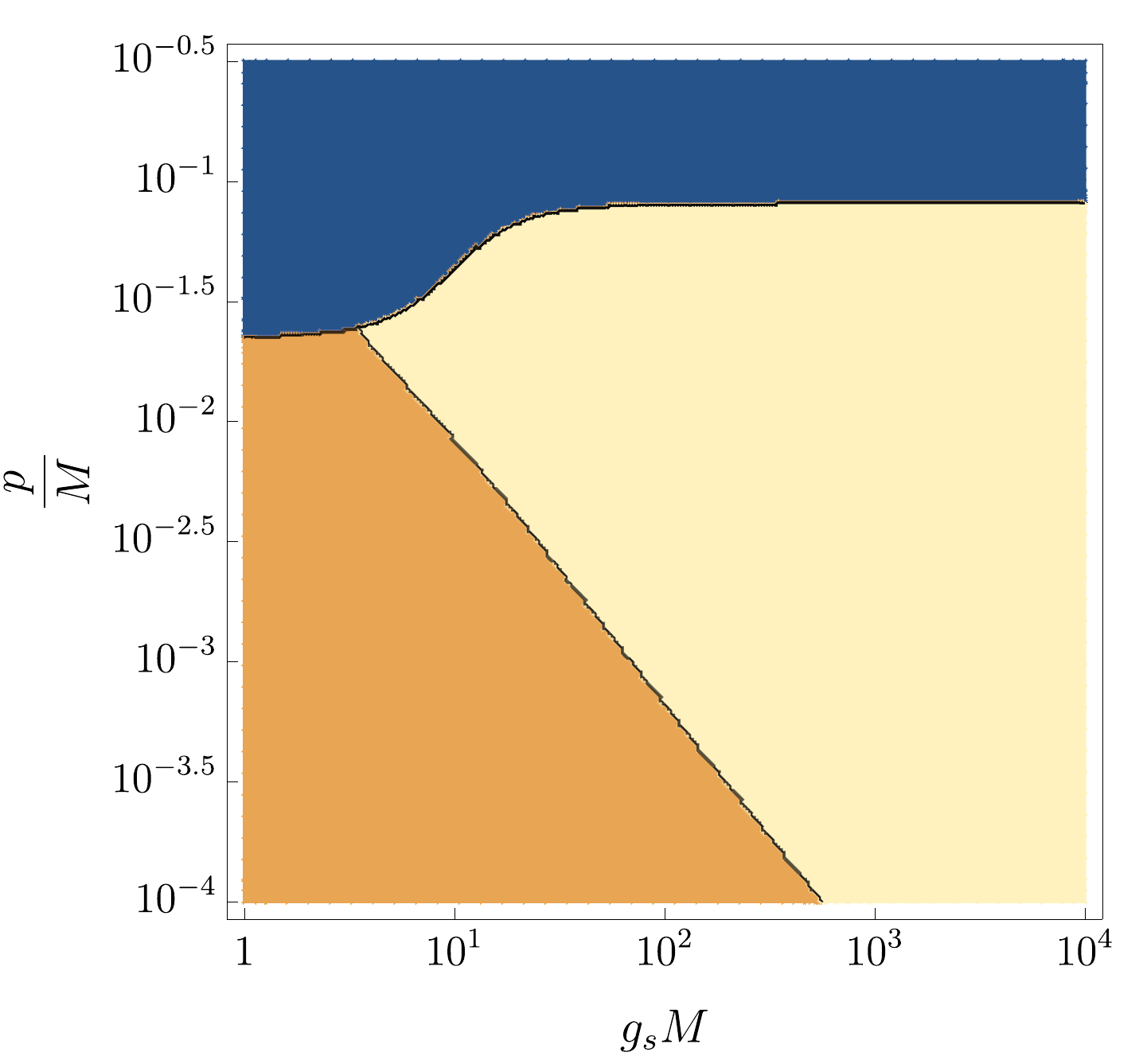}} 
    \subfigure[]{\includegraphics[width=0.47\textwidth]{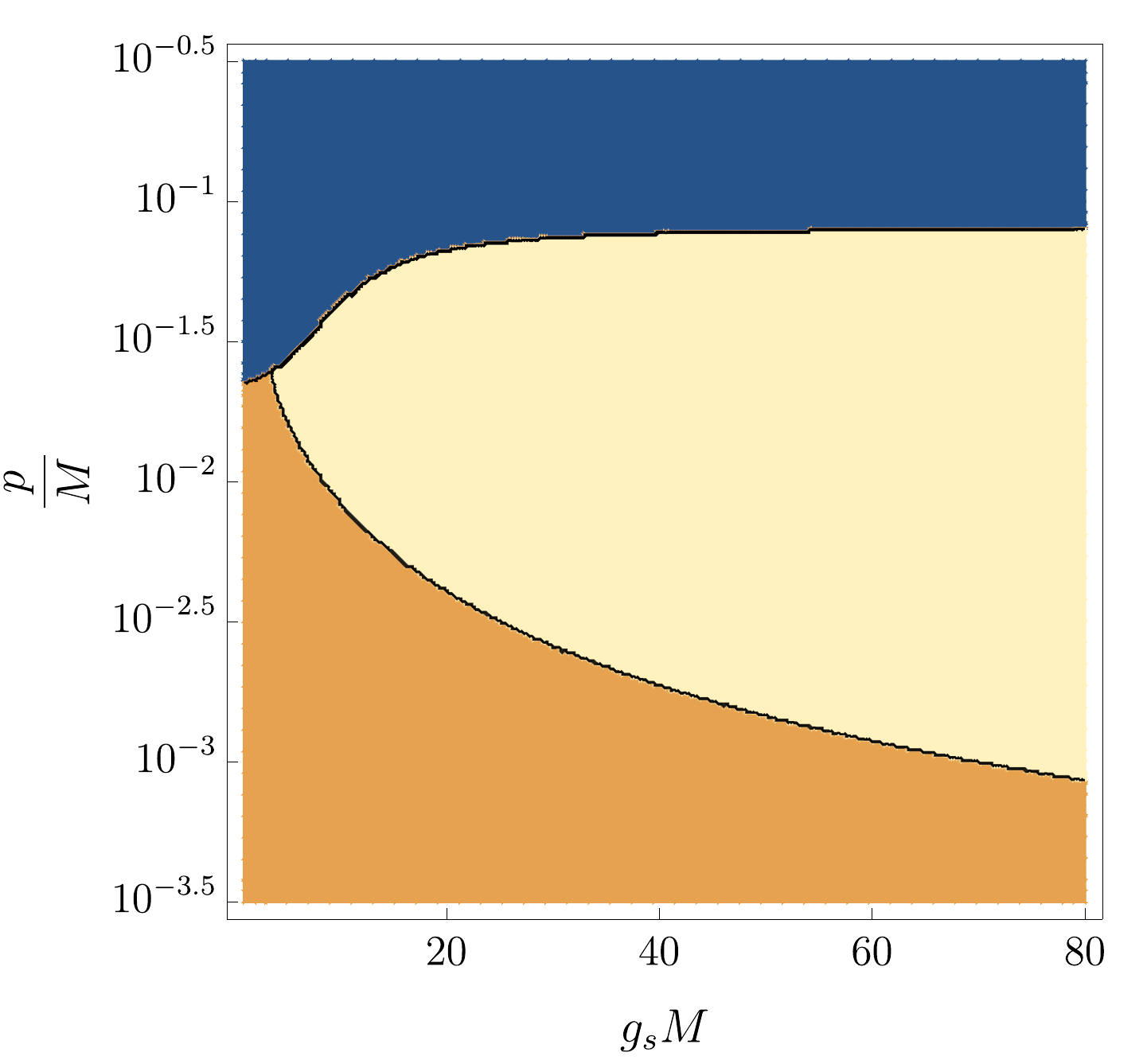}} 
    \caption{The yellow area shows the region in the $(g_sM,p/M)$-parameter space where a metastable minimum at positive energy exists. In the orange region, the minimum is at negative energy. In the blue region there is no metastable minimum, only an instability towards the supersymmetric minimum. In (a) a log-log-plot is shown for a wide range of parameters and (b) zooms into the region of smallest allowed values of $g_sM$.}
    \label{fig:poverm}
\end{figure}

The potential $V_\text{tot}$ is shown in Fig.~\ref{fig:totpotvarypoverM} for $g_sM=20$ and different values of $p/M$ and in Fig.~\ref{fig:totpotvarygsM} for $p/M=0.01$ and varying $g_sM$. As before, we see that depending on $p/M$ and $g_sM$ there exists a metastable minimum. 
A new feature is that the value of the potential at the metastable minimum, which we will call $V_\text{up}$, can be at negative, zero, or positive energy depending on the parameters. This can be summarized in the contour plot in Fig.~\ref{fig:poverm} where (a) is a log-log plot whereas (b) is linear on the $g_sM$ axis. The blue region describes the part of the parameter space where no metastable minimum exists, the yellow region describes the region with a metastable minimum at $V_\text{up}>0$, and the minimum is at negative energy in the orange region. As expected for large values of $g_sM$, the blue region is bounded from below by the KPV bound $p/M=0.08$.

Before discussing the crucial question of in which region of parameter space the potential \eqref{eq:vtot} can be trusted, let us comment on the main differences to the curvature corrected potential $V_\text{KPV}+V_\text{curv}$ calculated in \cite{Hebecker:2022zme}. As the highest order divergence in $1/\psi$ in \eqref{eq:vtot} has a positive prefactor coming from the correction $\Omega^4\mathcal{F}_2^2$, the potential diverges to $+\infty$ when $\psi\to0,\pi$. This facilitates the existence of a metastable minimum compared to \cite{Hebecker:2022zme} since also for very small values of $g_sM$, a minimum will always exist as long as $p/M$ is sufficiently small. 
Let us emphasize that in this regime, the radius of the NS5-brane is string size and all higher order $\alpha'$ corrections will become important. We will further elaborate on this below. As already discussed in \cite{Hebecker:2022zme}, we assume that summing up all the (unknown) higher order $\alpha'$ corrections, we find a smooth potential without any divergences. We then expect the fully corrected potential to have a finite value at $\psi=0$. It may be reasonable to expect that the fully corrected value of the potential at $\psi=0$ will be given by $p$ times the fully corrected potential of an anti-D3-brane. 

In the regime of large $g_sM$ both potentials yield similar results.

Let us now turn to the question of when we can trust the approximations that have led to this potential. A first condition is that $ g_s M \sin^2 \psi\sim R^2_\text{NS5}$ should be sufficiently large. The reason is that higher order $\alpha'$ couplings feature a suppression by $R^2_\text{NS5}$ or $R^2_{S^3}$. This applies to curvature and flux couplings. Thus, in order for higher order $\alpha'$ corrections to be negligible one should ensure that $\tilde{R}^2_\text{NS5}=g_sM\sin^2\psi$ is sufficiently large\footnote{Note that we do not take $R^2_\text{NS5} =b_0^4 g_sM\sin^2\psi$ as the expansion parameter as the powers of $b_0$ are different in some of the on-shell $\alpha'$ corrections.}. Caution is required for couplings involving $\mathcal{F}_2$ flux since $\mathcal{F}_2$ scales in the potential as
\begin{equation}
   \mathcal{F}_2 \sim \frac{1}{g_s \sin^2\psi}\left(\frac{p}{M}-\psi+\frac{\sin(2\psi)}{2}\right)\,,
   \label{eq:f2scaling}
\end{equation}
which is only small for moderately small $\psi$. 

This leads to a second condition specific to the terms involving $\mathcal{F}_2$ in the potential \eqref{eq:vtot}. As can be seen from \eqref{eq:f2scaling}, for small $\psi$, $\mathcal{F}_2$ scales like $\mathcal{F}_2\sim p/\tilde{R}_{NS5}^2$. But when $\psi\sim \mathcal{O}(1)$, \eqref{eq:f2scaling} implies $\mathcal{F}_2\sim - (M-p)/M$ which is $\mathcal{O}(1)$ in the regime where $p\ll M$. This is to be expected since with increasing $\psi$ the anti-D3-brane charge of the NS5 is neutralised by flux such that the NS5 describes $M-p$ D3-branes at the north pole. Hence, we lose control over $\mathcal{F}_2$ corrections close to the equator of the $S^3$ and on the north side of the equator $\psi>\pi/2$. This is not a serious issue since we anyways expect that the NS5 decays into the SUSY minimum at $\psi=\pi$ as soon as the NS5 slips over the equator. Note that it is not guaranteed that the SUSY minimum at $\psi=\pi$ has zero vacuum energy. The SUSY minimum has $M-p$ D3-branes whose $\alpha'$ corrected worldvolume action contributes to the potential as already discussed in \cite{Hebecker:2022zme}. This can lead the SUSY minimum to have negative energy, allowing the metastable minimum to decay even when the metastable minimum has zero or negative energy.

\begin{figure}[h]
    \centering
    \subfigure[]{\includegraphics[width=0.50\textwidth]{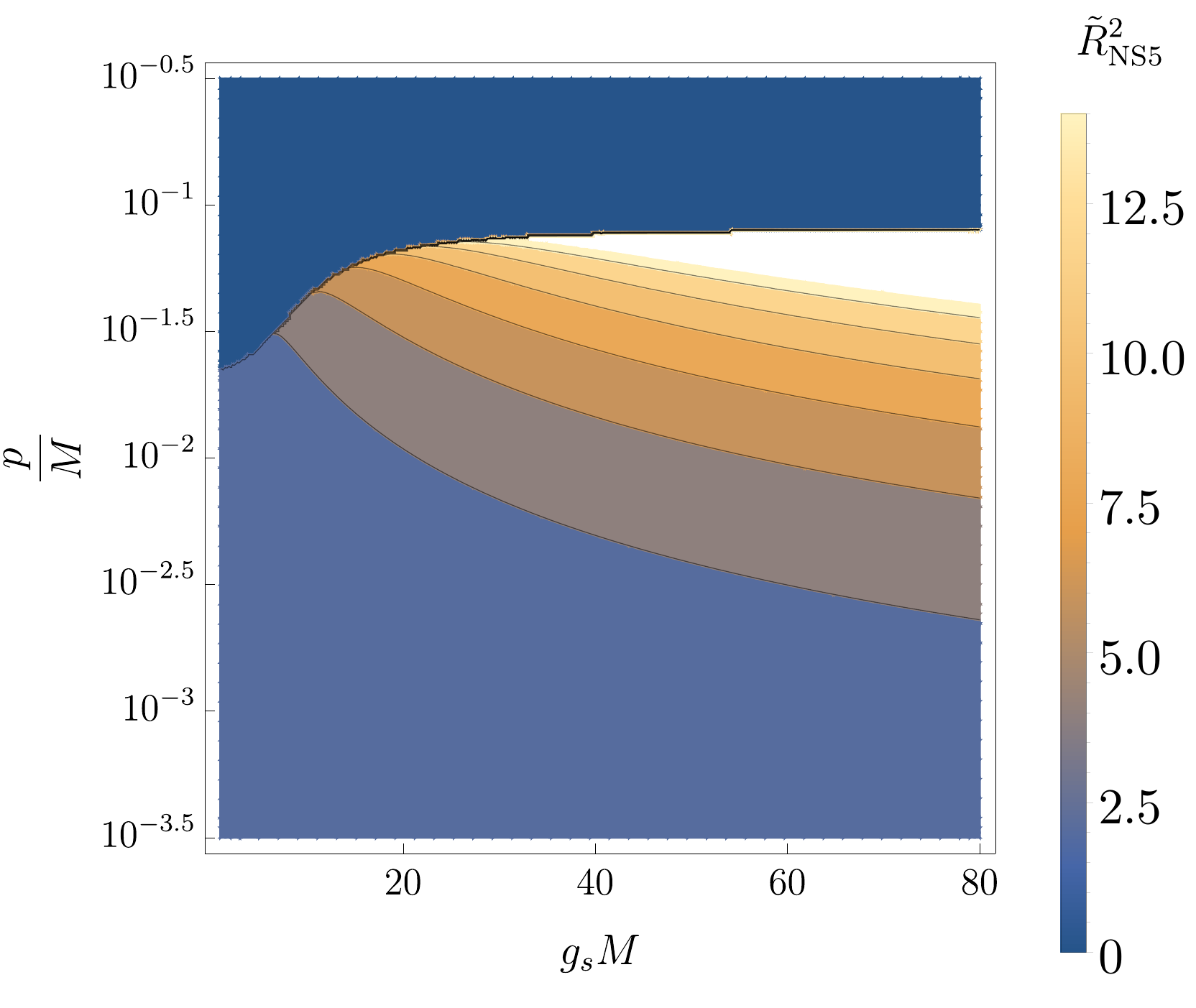}} 
    \subfigure[]{\includegraphics[width=0.49\textwidth]{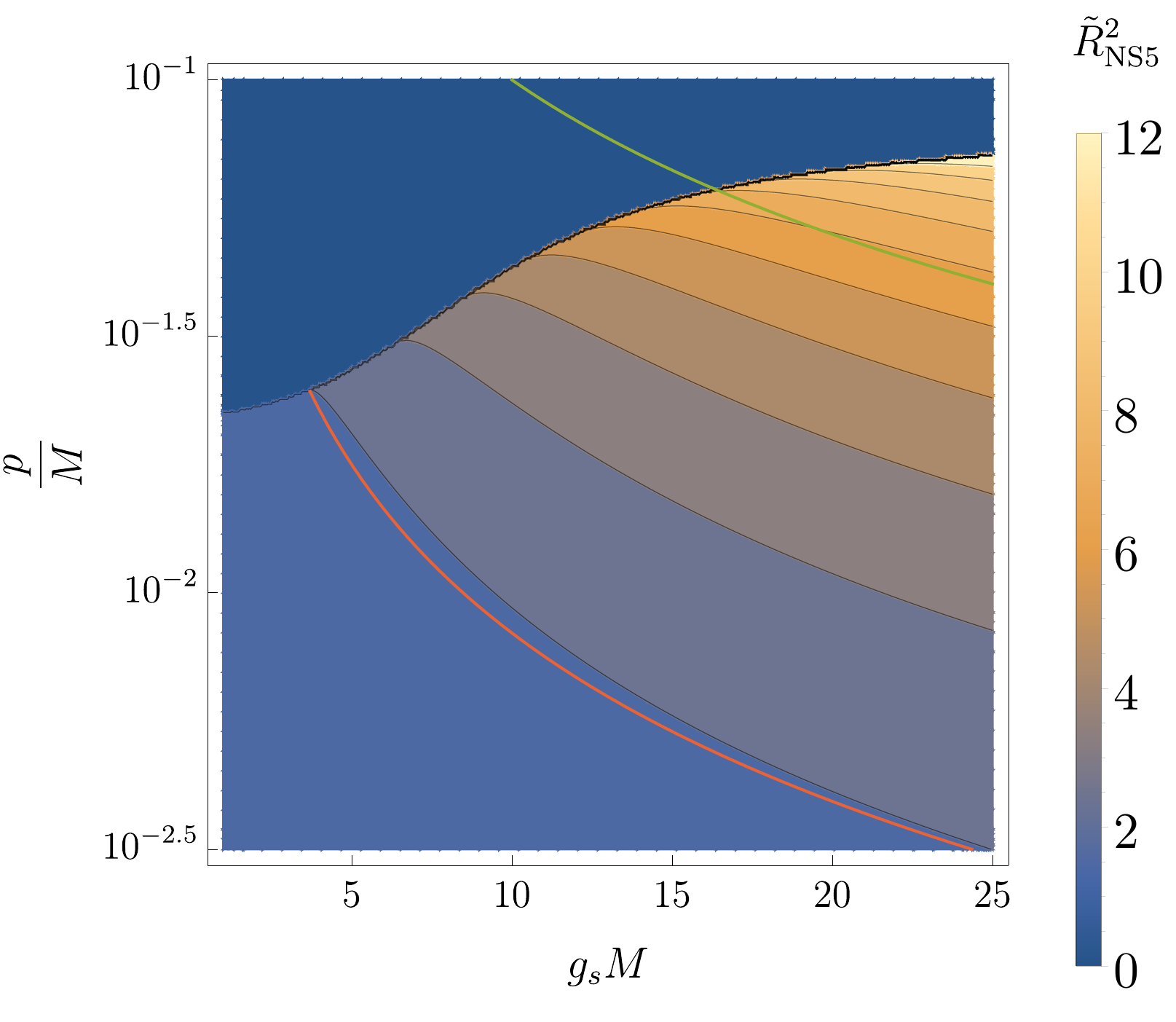}} 
    \caption{A similar contour plot to Fig.~\ref{fig:poverm}. The contours indicate lines of constant $\tilde{R}^2_\text{NS5}= g_sM\sin^2\psi$ in the $(g_sM,p/M)$ parameter space. Note that $\tilde{R}^2_\text{NS5}$ is proportional to the squared radius of the NS5-brane $R^2_\text{NS5}=b_0^4\tilde{R}^2_\text{NS5}$. The dark blue region is again the region where no metastable minimum exists. (a) depicts a wide range of parameters and (b) is a more detailed plot for smaller values of $g_sM$. In b) the contours are at integer values of $\Tilde{R}^2_\text{NS5}$ starting with $\Tilde{R}^2_\text{NS5}=1$ on the lowest contour. On the green line in (b), $g_s=1$ for $p=1$ and above the green line, $g_s>1$. The red line indicates the line where the minima of the potential are at zero energy.}
    \label{fig:rns5}
\end{figure}

Since in the interesting regime close to the south pole, $\tilde{R}^2_\text{NS5}$ is the crucial control parameter, Fig.~\ref{fig:rns5} depicts again the $(g_sM,p/M)$ parameter space but now with contours describing metastable minima of the potential of fixed $\tilde{R}^2_\text{NS5}$. Decreasing $g_sM$ also decreases the radius of the NS5. The regime where control over $\alpha'$ corrections is best is the large $g_sM$ and large $p/M$ regime. This again forces highly warped, deep throats as already observed in \cite{Junghans:2022exo,Gao:2022uop,Junghans:2022kxg,Hebecker:2022zme}. We discuss implications of this for phenomenology in Sect.~\ref{sec:deepthroat}.

In the cases where $V_\text{up}=0$, the radius of the NS5 is $\mathcal{O}(1)$ which marks the outermost edge of control as can be seen from the red line in Fig.~\ref{fig:rns5} (b), on which $V_\text{up}=0$ holds.
This is precisely the regime needed for the uplifting mechanism without deep throats which we will discuss further in Sect.~\ref{sec:newuplift}.

Note that in the regime of large $g_sM$ and $p/M$ one eventually enters the regime where $g_s>1$.
To stay in the weak coupling regime of type IIB string theory, we should ensure that
\begin{equation}
    \frac{p}{M} < \frac{p}{g_sM}\,,
\end{equation}
which drives one to smaller values of $R_\text{NS5}$. As also noted in \cite{Hebecker:2022zme}, this forbids for $p=1$ a large region of parameter space. The line $g_s=1$ for $p=1$ in the $(g_sM,p/M)$ parameter space is depicted in Fig.~\ref{fig:rns5} (b) in green. Hence, the parameter space above this line is forbidden when choosing $p=1$. With increasing $p$, the forbidden region shrinks such that the regime of large $g_sM$ and $p/M$ becomes accessible at weak coupling.

Besides the KPV decay channel, another proposed instability channel for antibranes at the tip of the KPV throat is the conifold transition \cite{Bena:2018fqc, Blumenhagen:2019qcg, Bena:2019sxm, Randall:2019ent,Scalisi:2020jal} (recently questioned in \cite{Lust:2022xoq}). To avoid a conifold instability for a single antibrane one imposes $\sqrt{g_s}M>6.8$. Note however that for the smallest values of $g_s$ and $M$ where KPV has a zero vacuum energy metastable vacuum, $\sqrt{g_s}M=12$, such that after $\alpha'$ corrections requiring the metastable vacuum to have a positive energy is a stronger constraint than that imposed by the conifold instability.

\section{Uplifting without a deep throat} \label{sec:newuplift}
We have seen from Fig.~\ref{fig:poverm} that with $\alpha'^2$ corrections accounted for, there are three types of metastable vacua allowed: those with a negative energy at sufficiently small $g_s M$ and $p/M$; those with a positive energy at sufficiently large $g_s M$ and $p/M$; and separating these a line of vacua with zero energy. The line of vacua with zero energy is approximately given by 
\begin{equation}
\label{eq:minkline}
    \frac{p}{M} \approx \frac{0.1029}{(g_s M)^{1.0909}}\,,
\end{equation}
and the metastable vacuum with zero energy exists for $g_s M \gtrsim 3.6$. The smallest throat with a zero energy metastable vacuum then has $p/M\approx 0.025$ and $g_s M \approx 3.6$, yielding for $p=1$ antibranes $M=40$ and $g_s = 0.09$.

As discussed in the Introduction, one of the  main applications of placing an anti-D3-brane at the tip of a warped throats a la KPV is as an uplifting mechanism. The idea is to start from a compactification with a scale-separated AdS vacuum with all moduli stabilized with potential $V_\text{AdS}$. One then introduces the antibrane as a source of positive potential energy $V_\text{up}$. One must be able to tune $V_{\text{up}}\approx |V_\text{AdS}|$ which in most cases requires a hierarchically small value $V_\text{up}$ to avoid a runaway. The existence of the line of metastable vacua with zero energy in the $(g_sM,p/M)$ parameter space now provides a novel uplifting mechanism.

Trusting the potential in the regime where vacua with zero energy exist, one can clearly obtain a metastable vacuum with a hierarchically small positive energy by fine-tuning $g_s$ to be arbitrarily close to the line \eqref{eq:minkline} where $V_\text{up}=0$. For this to work one must of course have a sufficiently densely spaced in $g_s$ discretuum of vacua in the IIB landscape. This mechanism can be seen in action by tuning close to the $g_sM =10$ line in Fig.~\ref{fig:totpotvarygsM}.

One may object to this uplifting mechanism on several grounds.

First, one must glue the KS throat into a compact geometry. The resulting geometry will not perfectly match the KS geometry. It seems reasonable that if the modification to the geometry is small the leading effect of this modification will be to alter the coefficients $c_i$ appearing in \eqref{eq:vtot} by a small amount. This is not deadly to the uplifting mechanism we propose so long as the corrections to the $c_i$ are sufficiently small that a line of $V_\text{up}=0$ continues to exist. However, one may worry that the geometry at the tip is more seriously deformed and the zero energy vacua are lost. One method to achieve a modicum of safety is to remember that the warping at the tip is given by $\exp(-8\pi K/3 g_sM)$ and a deep warped throat to some extent decouples the tip of the throat from the bulk geometry (see e.g.~\cite{Kachru:2007xp,Berg:2010ha}). When $K$ is too small there is really no decoupling at all. However, one could demand $8 \pi K > 3 g_sM$ to ensure there is some amount of decoupling between tip and bulk while at the same time not making $K$ too large to avoid problems such as the singular bulk problem in KKLT or similar issues in LVS that arise when one needs a very deep throat when one attempts to make the uplifting contribution to the potential hierarchically small through only the warping at the tip. In fact, one could combine warping and our uplifting mechanism in a mutually beneficial manner: One achieves part of the hierarchic suppression in $V_\text{up}$ through warping but stops before one runs into issues with the throat being too big. The remainder of the hierarchy is then achieved through our mechanism; where the required fine-tuning in $g_s$ is now reduced by several orders of magnitude; making the new mechanism in turn easier to implement as one requires a less dense discretuum in $g_s$.

In fact for the smallest zero vacuum energy throat $g_s M \approx 3.6$ the constraint to have some warping at the tip is that $K> 3 g_s M / 8\pi \approx 0.43$ which is trivially satisfied and already for $K=1$ one has a not insubstantial amount of warping at the tip $\exp(-2.33) \approx 10^{-1}$. By choosing $K=2,\,3$ one can already obtain several orders of magnitude of warping while maintaining a reasonable D3-tadpole for the throat.

Second, as can be seen from Fig.~\ref{fig:rns5}, the line of zero energy metastable minima occurs when $R_\text{NS5} \approx 1$. By attempting to fine-tune ourselves arbitrarily close to this line, we come arbitrarily close to the boundary of control and it is highly dubious how reliably our results are as all orders in $\alpha'$ become important. We can only hope that our results give a hint of the behaviour in this regime but clearly the abelian fivebrane picture is not suitable to analyse this regime. Ideally, one would analyse this regime in another picture such as holographically or as a nonabelian D3-brane stack. In fact, $g_s M \ll 1$ is the regime where the holographic picture is perturbatively controlled. If one could establish holographically that there exists a metastable AdS minimum somewhere in the regime $g_s M \ll 1$, $p/M \lesssim 0.025$, then this would qualitatively confirm the picture in Fig.~\ref{fig:poverm} as at large $g_s M$ we know we have controlled metastable vacua with positive energy and by continuity one then expects vacua with zero energy in some intermediate regime.

Third, we have performed our analysis in a probe approximation and have not considered the backreaction of the fivebrane. Backreaction generally was commented on in \cite{Hebecker:2022zme} and we will not rehash those general remarks here. Let us make two specific points. A first point is that \cite{Cohen-Maldonado:2015ssa} have noted that backreaction may result in the metastable minimum being at parametrically larger $R_\text{NS5}$ than the probe approximation suggests. In this case, the novel uplifting mechanism we propose may occur at better controlled $R^2_\text{NS5}$ than our results suggest and it would be interesting to investigate this interplay. A second point is specific to KKLT. In recent work \cite{Bena:2022ive} have noted that the branes at the tip of the throat source $(0,3)$ three-form flux which in turn can give a mass to gauginos on D7-branes. If in KKLT one stabilized the Kahler moduli using D7-branes, these gauginos are required to condense for moduli stabilization and so must have a mass $m_\lambda$ below the confinement scale $\Lambda_c$. It was shown by \cite{Bena:2022ive} that this implies
\begin{equation}
\label{eq:gauginoback}
    \frac{m_\lambda}{\Lambda_c} \approx 10^2 \frac{1}{M^{1/4}} \frac{p}{M}\frac{1}{(g_s M)^{5/4}} f(2 \pi K / g_s M) \ll 1\,,
\end{equation}
with $f(x)= x^{5/4} \exp{(-19 x / 9)}$. The claim is that for small throats this constraint will not be satisfied and so one must demand a sufficiently large throat for the uplift. Note however that our novel uplifting mechanism always satisfies these constraints, even for the smallest throats where the mechanism operates. The function $f(x)$ is bounded from above by $0.14881$ and the smallest throats allowed by our mechanism obey $p/M \approx 0.025$ and $g_s M \approx 3.6$ such that the bound becomes
\begin{equation}
    \frac{m_\lambda}{\Lambda} \approx 10^{-1} \frac{1}{M^{1/4}} \ll 1\,,
\end{equation}
which is always satisfied by the quantization of $M$. It then follows that in KPV if the anti-D3-branes polarize into an NS5-brane the loss of control over $\alpha'$ corrections becomes important before the bound \eqref{eq:gauginoback} does.

\section{Deep throat phenomenology} \label{sec:deepthroat}
In the previous section we discussed a novel uplifting mechanism. In this section we will for the sake of comparison instead consider the traditional mechanism where one also uplifts via $p$ anti-D3-branes at the tip of a warped throat. However, in the traditional mechanism the uplifting condition $|V_\text{AdS}| \approx |V_\text{up}|$ is achieved solely through a large amount of warping at the tip of the throat and one demands that $\alpha'$ corrections to $V_\text{up}$ are sufficiently small for control. This leads to the requirement that one has a large warped throat, which is a severely constraining condition on the possibility to achieve an uplift.

For KKLT with large throats there is the singular bulk problem \cite{Freivogel:2008wm,Carta:2019rhx, Gao:2020xqh} which generically appears deadly (see however \cite{Carta:2021lqg}). Thus we will focus on the LVS where large warping is required in order to control the most dangerous corrections to the LVS scalar potential \cite{Junghans:2022exo, Gao:2022fdi, Junghans:2022kxg,Hebecker:2022zme}. One such constraint is the Parametric Tadpole Constraint (PTC) \cite{Gao:2022fdi} which quantifies how large the negative contribution to the D3-tadpole, $Q_3$, needs to be for control. In App.~\ref{sec:TadpoleAppendix} we review the PTC in some detail and comment on how dangerous we expect loop corrections to the scalar potential to be to explain why these are not accounted for in the PTC in contrast to \cite{Junghans:2022exo,Junghans:2022kxg}.

It was already observed in \cite{Hebecker:2022zme} that the fundamental parameters of $\alpha'$ corrected KPV are $p/M$ and $g_sM$. We have seen from Fig.~\ref{fig:rns5} that increasing $g_sM$ by keeping $p/M$ as large as possible (required that $g_s<1$) increases the control over $\alpha'$ corrections. The reason being that our parameter of control, the radius of the $S^2$ wrapped by the NS5-brane, increases. Clearly, if we demand that we have significant control, then we must uplift using a deep warped throat. 

Hence $g_sM$ is the main control parameter and it is therefore useful to reformulate the PTC such that it takes $c_e= g_sM$ As an input control parameter. This version of the PTC is derived in App.~\ref{sec:gsmptc} with the result that the total negative D3-tadpole of the compactification geometry $Q_3$ must obey the bound
\begin{equation}
    \hspace*{-.3cm}-Q_3 > N_{\{c_N,c_e\}} = \frac{2^{8/3}}{\pi} \frac{\kappa_s^{2/3} \, c_e^2 }{\xi^{2/3} \, a_s} \left( \mathcal{W}_{-1}(y)\right)^2 ,\,\,  y= - \frac{3^{2/5}\,a_0^{1/4}}{8\,\,2^{23/40}\, 5^{5/8}\, \pi^{1/40}} \, \frac{\kappa_s^{1/6}\, \xi^{1/12}}{p^{1/4}\, a_s^{1/4}\, c_N^{5/8}\, c_e^{1/4} }
    \,,
    \label{eq:gsmptc}   
\end{equation}
in order for the traditional de Sitter uplift with a large warped throat to be possible. Here $\kappa_s$ is related to the triple self intersection number of the divisor associated to the LVS blow-up cycle, $\xi=0.6\chi/(2\pi)^3$ with the Euler number $\chi$ of the Calabi-Yau on which is compactified, $a_s$ stems from the nonperturbative corrections to the superpotential responsible for stabilising the Kahler moduli\footnote{If the nonpertubative corrections are Euclidean D3-branes, one has $a_s=2\pi$. For gaugino condensation on a stack of D7-branes, $a_s$ depends on the gauge group, for instance $a_s=\pi/3$ for $SO(8)$.},  $\mathcal{W}_{-1}(x)$ is the -1 branch of the Lambert $\mathcal{W}$ function, and $a_0\approx 0.71805$. The control parameter $c_N$ quantifies the control over a correction due to a varying warp factor in the bulk of the Calabi-Yau.

Since the topological quantities $(a_s,\xi,\kappa_s)$ enter polynomially in \eqref{eq:gsmptc}, we can see that they are an essential ingredient to find suitable models where the PTC is weakest but still satisfies the desired amount of control chosen by $c_e$ and $c_N$. This is also emphasized in \cite{Junghans:2022exo,Junghans:2022kxg}. The quantities $a_s$ and $\xi$ should be chosen as large as possible and $\kappa_s$ as small as possible to minimize $(-Q_3)_\text{min}$.

Combining the results of Sect.~\ref{sec:fluxcorretions} with the PTC will yield a bound on the minimal negative contribution to the D3-tadpole that respects the constraints coming from the $\alpha'$ corrected KPV potential.

To obtain the minimal tadpole for some examples we proceed as follows. First, we choose some reasonable numbers for $a_s$ and $\kappa_s$, namely $a_s=2\pi$, $\kappa_s=0.1$.
Then, to minimize $(-Q_3)_\text{min}$, we take $p/M$ as large as possible (such that $M$ is as small as possible for fixed $p$) for a given value of $g_sM$. This can be read off from Fig.~\ref{fig:rns5}. This will fix $p$ (which we may take as small as possible still compatible with $g_s<1$) and $g_s$. 

These parameters together with some value of $c_N$ determine $\xi$ via \eqref{ceofcm} with the PTC \eqref{eq:gsmptc} plugged in for $N$. One obtains 
\begin{equation}
    \xi^{-2/3} \mathcal{W}_{-1} (y) = \frac{3}{2^{11/3}}\, \frac{a_s}{\kappa_s^{2/3}} \, \frac{M}{p} \, \frac{p}{c_e}\,,
\end{equation}
which can be solved numerically for $\xi$. With this $\xi$ and the other input parameters, the minimal D3-tadpole $(-Q_3)_\text{min}$ is readily obtained using \eqref{eq:gsmptc}. 

The minimal choice of $g_sM$ depends on the readers notion of control. Each point in the $(g_sM,p/M)$ parameter space corresponds to a specific radius of the NS5 which determines the amount of control over higher order $\alpha'$ corrections. For this reason, we list $(-Q_3)_\text{min}$ for multiple values of $\tilde{R}_\text{NS5}$. The results are summarized in Tab.~\ref{tab:tadpole}.

\begin{table}[!h]\centering
	\caption{The minimal value of the required negative contribution to the D3-tadpole $(Q_3)_\text{min}$ for $a_s=2\pi$ and $\kappa_s =0.1$ for different values of the control parameter $\tilde{R}_\text{NS5}^2=g_sM\sin^2\psi$. The choice of $g_sM$ and $p/M$ for a given value of $\tilde{R}_\text{NS5}^2$ is such that the tadpole is minimal.}
	\vspace{.3cm}
	\label{tab:tadpole}
		\begin{tabular}{cccccccccccccc}
		\toprule
		    \multicolumn{5}{c}{input parameters} && &  & &\multicolumn{2}{c}{$c_N=5$} & & \multicolumn{2}{c}{$c_N=100$}\\
		    \cmidrule(lr){10-11}  \cmidrule(lr){13-14}
			 $\tilde{R}^2_\text{NS5}$&$p$&&$g_s M$&$p/M$ &&$g_s$&$g_s M^2$& & $\chi$ & $(-Q_3)_\text{min}$ && $\chi$ & $(-Q_3)_\text{min}$ \\
			\hline
			\midrule
			1&1 &&3.8 &0.025 & &0.095 & 152&  &14 &560& &20 & 715 \\
            \midrule
			2.2&1 && 7&0.0323 & &0.226 &217 &  & 52&803& &75 & 1015\\
			\midrule
			3&1 && 9.5& 0.0385& &0.365 &247 &  &108 &913& &155 & 1156\\
			\midrule
			5.2&1 &&13 & 0.0526&  &0.684 & 247& & 278& 913 && 397&1156\\
           \midrule
			9&2 && 20.7&0.0667 &  & 0.69& 310& &299 &2387 &&422 &3000\\
			\bottomrule
	\end{tabular}
\end{table}

Comparing these numbers with \cite{Hebecker:2022zme} we observe that the $\alpha'$ corrections calculated in this work lead to higher values of $(-Q_3)_\text{min}$. The main reason being that at small values of $g_sM$, also $p/M$ has to decrease such that a metastable minimum remains. This finally leads to larger values of $M$ (and smaller values of $g_s$) such that the volume of the Calabi-Yau and the tadpole in the throat increase. 
In contrast to \cite{Hebecker:2022zme} where the upper bound on $p/M$ at smallish $g_sM$ is always given by the KPV bound $p/M<0.08$. Interestingly, as can be seen from line four and five in Tab.~\ref{tab:tadpole}, increasing the control over $\alpha'$ corrections does not necessarily result in a higher tadpole. The reason for this is again that higher values of $p/M$ are allowed when $g_sM$ is increased. Let us also emphasize that the control parameter $\tilde{R}_\text{NS5}$ increases only slowly at small $g_sM$. Thus, a slight in- or decrease in control can already have a significant impact on the resulting tadpole.

In total, the results point towards large tadpoles already at small values of $g_sM$ in order to have control over $\alpha'$ corrections. This challenges LVS model-building to find Calabi-Yaus with large tadpole.

Considering the strong constraints on $(-Q_3)_\text{min}$ in the traditional uplift with large warped throats, it becomes very attractive to study the new way of uplifting discussed in Sect.~\ref{sec:newuplift}. There, the PTC is not applicable since $V_\text{up}$ can be tuned exponentially small without large warping. Instead, we saw that this novel uplifting mechanism already works for $M=40$, $K=1$ such that for the novel uplifting mechanism one has the much more reasonable constraint $-Q_3>KM=40$ for the negative D3-tadpole of the compactification geometry.

\section{The S-dual KS set-up}
\label{sec:sdks}

In this paper we have focused on the Klebanov-Strassler throat. One may instead consider an anti-D3 at the tip of the S-dual to the Klebanov-Strassler throat (SDKS) \cite{Gautason:2016cyp}. 
The advantage to this is that in the SDKS set-up one works with D5-branes for which the $\alpha'$ corrections to the fivebrane worldvolume action we consider are explicitly known, while for the NS5-brane in KS we had to infer the analogous $\alpha'$ corrections by the argument described in App.~\ref{sec:SdualAppendix} which relies on the non-trivial assumptions 1)-4) given in the Appendix.

In the SDKS geometry, there are $K$ units of $H_3$ flux on the $S^3$ at the tip of the throat and $M$ units of $F_3$ flux on the B-cycle. In this S-dual set-up $p$ anti-D3-branes will puff up into a D5-brane wrapping an $S^2$ at the tip of the throat, rather than an NS5-brane. The radius of the tip is $\sim \sqrt{K}$ in SDKS rather than $\sim \sqrt{g_s M}$ in KS. It is straightforward to compute the $\alpha'$ corrected worldvolume potential for the D5-brane at the tip of the SDKS throat by going through the analysis of Sect.~\ref{sec:fluxcorretions} again but now with the SDKS geometry and the action of the D5-brane. One obtains \eqref{eq:vtot} but with $g_sM$ substituted by $K$ and $p/M$ substituted by $p/K$ for the potential of the D5-brane. The analysis of Sect.~\ref{sec:fluxcorretions} then also goes through for the D5, with these substitutions. For instance, Fig.~\ref{fig:poverm} and Fig.~\ref{fig:rns5} also show the different regimes of the D5-brane after substituting the $g_sM$ axis by a $K$ axis. and the $p/M$ axis by a $p/K$ axis.

The issue now is that $K$ and $p$ are both integer and so we cannot achieve arbitrary points Fig.~\ref{fig:poverm} and Fig.~\ref{fig:rns5}. Crucially, to have a metastable vacuum we require at the very least $p/K < 0.08$, which requires $K > 12.5  p$. Since then at the very least $K=13$ it is not possible to achieve a small throat in the SDKS set-up, preventing the use of a small throat uplift.

One may also repeat the analysis of Sect.~\ref{sec:deepthroat} for a traditional uplift with a large warped throat in the LVS for the SDKS set-up.

To obtain the corresponding formulas, replace $g_sM\to K$ and $K \to g_s M$ everywhere. One finds
\begin{equation}
    V_\text{uplift} =\frac{\left( 3^2\,\pi^3\, 2^{22/3} \right)^{1/5}}{a_0} \frac{g_s}{K^2\mathcal{V}^{4/3}}\text{e}^{-\frac{8\pi g_s N}{3 K^2}}\,.
\end{equation}
The derivation of the PTC leads precisely to \eqref{ptc2} but with the replacement $c_e\to K$. The same conclusion that for an uplift with a large warped throat one requires a very large negative D3-tadpole in the compactification geometry then also holds when using SDKS to uplift. However, in SDKS unlike in KS there is no hope of achieving an uplift in a small throat using the novel uplifting mechanism.

\section{Conclusions and outlook}
\label{sec:conclusions}

The main result of this paper is to compute as completely as we were able the $\alpha'$ corrected potential for the NS5-brane at the tip of a warped throat.
As a necessary, intermediate step we infer the $\alpha'$ corrected NS5-brane action from the corresponding $\alpha'$ corrected D-brane action. The underlying argument is based on several nontrivial assumptions discussed in App.~\ref{sec:SdualAppendix}. Therefore, the $\alpha'$ corrected potential for the NS5-brane is also based on these assumptions. This potential is given by $V_{\text{tot}}$ in \eqref{eq:vtot}. The resulting potential matches the results of KPV in the regime where $\alpha'$ corrections are small, but several interesting new features appear when $\alpha'$ corrections become important.

At large $g_s M$, the condition to have a metastable KPV vacuum is $p/M\lesssim 0.08$ as in KPV. For small $g_s M$, this  bound becomes  stronger as can been seen in Fig.~\ref{fig:poverm}, getting as strong as $p/M\lesssim 0.025$ in the small $g_s M$ limit.

The $\alpha'$ corrections generically lower the potential of the metastable vacuum compared to the tree-level result. Once the corrections become very strong, the metastable vacuum can even obtain a negative energy as seen in Fig.~\ref{fig:poverm}. There exists then a line of metastable vacua with zero energy. This presents the possibility of a novel uplifting mechanism: By keeping the vacuum energy of the metastable minimum positive but tuning $g_s$ to get arbitrarily close to the line of zero-energy vacua one can obtain an arbitrarily small uplifting potential $V_{\text{up}}$ in a small throat without needing to rely on warping. This permits an uplift for a throat with a D3-tadpole contribution as low as $N=40$. 

This is in contrast to the traditional anti-D3 uplift using exponential warping. The requirement to have a sufficiently large warped throat for the standard uplifting mechanism is generically impossible to satisfy in KKLT due to the singular bulk problem. In the LVS one requires compactification geometries with $\mathcal{O}(10^3)$ negative D3-tadpole which are difficult to obtain. The fact that these issues are avoided in our new small throat uplifting mechanism makes the new uplifting mechanism we propose in our opinion very promising.

The main issue with the small throat uplift is that it relies on $\alpha'$ corrections to achieve a very small $V_{\text{up}}$. As a result, at the metastable minimum the $\alpha'$ corrections to the potential are of the same order of magnitude as the tree-level potential. By the very nature of our set-up, this uplift then drives us near the boundary of control over the perturbation series of $\alpha'$ corrections as seen in Fig.~\ref{fig:rns5}.

One way to deal with this issue is to continue computing higher order $\alpha'$ corrections until one is confident that one has control despite being near (but not past) the boundary where the perturbation series in $\alpha'$ breaks down. In a way this is what we have initiated here. In \cite{Hebecker:2022zme} it was suggested that the new uplifting mechanism might be possible, but this could not explicitly be checked as not taking into account $\alpha'$ flux corrections led to some clearly unphysical behaviour and we had to speculate how this unphysicality would resolve. The explicit computations we have done here confirm the qualitative behaviour guessed at in \cite{Hebecker:2022zme}. Unfortunately, to compute even higher order $\alpha'$ corrections to the fivebrane worldvolume seems a rather grueling task, especially as we have accounted for all $\alpha'$ corrections to the worldvolume whose explicit form we were able to find in the literature. Any further terms one then first has to derive the form of and then compute in the KS throat background.

However, there is also good news when it comes to control.
First, the abelian NS5-brane analysis is just one perspective on the metastable KPV vacuum. One may also perform the analysis from the perspective of either a nonabelian stack of anti-D3-branes or a holographic perspective. In fact, the regime $g_s M \ll 1$ where the $\alpha'$ perturbation theory breaks down for the NS5-brane is precisely the regime where the holographic picture is under good perturbative control. If one could holographically establish that there exists a metastable vacuum with negative or zero energy for $g_s M \lesssim 1$, $p/M \lesssim 0.025$ as predicted in Fig.~\ref{fig:poverm}, then under the assumption that the potential at the metastable minimum is continuous in $g_s M$ and $p/M$, it is guaranteed that there exist metastable vacua with zero energy and our new uplifting mechanism works. Second, the issues with control in our analysis seems isolated to the tip of the throat, which in compactifications one can isolate from the rest of the geometry by an intermediate amount of warping. This is in contrast to uplifts with a large warped throat which can strongly affect and destroy the entire compactification geometry as for instance with the singular bulk problem in KKLT \cite{Freivogel:2008wm,Carta:2019rhx, Gao:2020xqh}.

We believe that our work shows very encouraging evidence that a de Sitter uplift using $\alpha'$ corrections for anti-D3-branes in small warped throats is possible. This avoids the issues with uplifts using large warped throats. Questions concerning the amount of control in our set-up remain. Many future directions to analyse our set-up are open and we hope to continue this story in the future.
\section*{Acknowledgements}
We thank Arthur Hebecker for comments on a draft of this manuscript and valuable discussion. We thank Mohammad Garousi and Daniel Junghans for valuable discussion. 
We thank an anonymous referee for valuable comments on an earlier draft of this paper.
This work was supported by the Graduiertenkolleg ‘Particle physics beyond the Standard Model’ (GRK 1940).

\appendix

\section{Flux corrections to branes and their evaluation at the tip of the throat}
\label{sec:fluxcorrections}

In this appendix, we evaluate the $\alpha'$ corrections calculated in \cite{Garousi:2009dj,Garousi:2010ki,Garousi:2010rn,Garousi:2011ut,Robbins:2014ara,Garousi:2014oya,Jalali:2015xca,Garousi:2015mdg,Jalali:2016xtv,BabaeiVelni:2016srs,Garousi:2022rcv} at the tip of the throat. 
To do so, we need the KS fluxes close to the tip of the throat. They are given by\footnote{We only display terms which are relevant to our calculations.}
\begin{align}
    \label{eq:b2}
    B_2  \supset &\,\frac{g_s M \alpha' \tau}{6} g^3 \wedge g^4+\mathcal{O}(\tau^2)\,, \\
    \label{eq:h3}
    H_3 \supset &\,\frac{g_s M \alpha' }{6}  \dd\tau\wedge g^3\wedge g^4 + \frac{g_s M \alpha' \tau}{12} g^5\wedge\left(g^1\wedge g^3 + g^2\wedge g^4\right)+\mathcal{O}(\tau^2)\,,\\
    \label{eq:f3}
    F_3  \supset& \,\frac{M\alpha'}{2} g^5\wedge g^3 \wedge g^4+ \frac{M \alpha' \tau}{12}\dd\tau\wedge\left(g^1\wedge g^3 + g^2\wedge g^4\right)+ \mathcal{O}(\tau^2)\,,\\
    \label{eq:f5}
    \tilde{F}_5  \supset& \left(\frac{\tau}{3^{4/3} g_s^3 M^2 a_0^2}+\mathcal{O}(\tau^3)\right) \dd x^0 \wedge \dd x^1 \wedge \dd x^2 \wedge \dd x^3 \wedge \dd\tau+\mathcal{O}(\tau^2)\,,\\
    \label{eq:h7}
    H_7  \supset& - \frac{1}{2^{5/3} a_0 g_s^3 M}\dd x^0 \wedge \dd x^1 \wedge \dd x^2 \wedge \dd x^3 \wedge g^3 \wedge g^4 \wedge g^5+\nonumber\\
    &\frac{\tau }{12\,2^{5/3} a_0 g_s^3 M}\dd x^0 \wedge \dd x^1 \wedge \dd x^2 \wedge \dd x^3 \wedge\dd\tau\wedge\left(g^1\wedge g^3+ g^2\wedge g^4\right) +\mathcal{O}(\tau^2) \,,
\end{align}
where the one forms $g^{1\dots5}$ (see for instance \cite{Herzog:2001xk}) parametrize the $T^{1,1}$.

Before explicitly evaluating $\alpha'$ corrections to the worldvolume action of a fivebrane wrapping an $S^2$ at the tip of the throat, we list some important properties of the fluxes that will make many corrections vanish.
\begin{enumerate}
    \item[a)] There are only a few components of the fluxes that are non-vanishing at the tip. Hence, a wrong number of tangential and/or normal indices will make terms vanish.
    \item[b)] $F_3$, $C_2$ and $F_2$ are covariantly constant along the $S^2$. That means that\footnote{We work in conventions where $(\alpha,\beta,\gamma,\cdots)$ are indices tangent to the brane and $(a,b,c,\cdots)$ are indices normal to the brane.} $\nabla_\alpha \mathcal{F}_{2\,\alpha\beta } =0$ and $\nabla_\alpha F_{3\, \beta\gamma a}=0$. 
    \item[c)] The covariant derivative of $F_3$ with respect to a normal index is always zero at the tip except for the term $\nabla_\psi F_{3\, \theta\varphi\psi}$ where $(\theta,\varphi)$ parametrize the non-shrinking $S^2$ at the tip and $\psi$ the additional direction inside the $S^3$. 
    \item[d)] If covariant derivatives with respect to a normal index of form fields appear then components linear in $\tau$ can be non-vanishing at the tip if the terms in the action have the correct index structure. 
\end{enumerate}

\subsection{Flux corrections that vanish for the KS set up}

Let us start with flux corrections to the DBI action. There are $\alpha'^2$ corrections involving only covariant derivatives of $H_3$ \cite{Garousi:2009dj,Garousi:2014oya}
\begin{equation}
\label{eq:h3squared}
\begin{split}
    S_{\text{D$p$},\text{DBI}} \supset \frac{\mu_p}{g_s}  \frac{\pi^2\alpha'^2}{48} \int \dd^{p+1}\xi &\sqrt{-(g+2\pi\alpha'\mathcal{F}_2)} \biggl[  -\frac{1}{6} \nabla_\alpha H_{abc} \nabla^\alpha H^{abc}\\ & - \frac{1}{3}  \nabla_a H_{\alpha\beta\gamma} \nabla^a H^{\alpha\beta\gamma} 
    +\frac{1}{2} \nabla_\alpha H_{\beta\gamma a} \nabla^\alpha H^{\beta\gamma a} \biggr]\,,
\end{split}
\end{equation}
where $2\pi\alpha'\mathcal{F}_{2\,\alpha\beta}=2\pi \alpha' F_{2\,\alpha\beta}+B_{2\,\alpha\beta}$ for D$p$-branes and $2\pi\alpha'\mathcal{F}_{2\,\alpha\beta} \to 2\pi g_s \mathcal{F}_{2\,\alpha\beta}=2\pi g_s \alpha' F_{2\,\alpha\beta} - g_s C_{2\,\alpha\beta}$ for the NS5-brane. Evaluating \eqref{eq:h3squared} at the tip of the throat for an NS5-brane (where $H_3\to -F_3$ and the correct $g_s$ scaling can be inferred from analogous considerations as in Sect.~\ref{sec:fluxcorretions}) gives zero due to property b) and since the term $\nabla_\psi F_{3\, \theta\varphi\psi}$ does not show up (see property c)). 

Next, in \cite{Jalali:2015xca,Jalali:2016xtv} corrections including $\mathcal{F}_2$ are computed\footnote{We abbreviate $\mathcal{F}_2$ with $\mathcal{F}$ to not clutter notation.}:
\begin{equation}
\label{eq:furthermixedterms}
    \begin{split}
        S_{\text{D$p$},\text{DBI}} \supset & \frac{\pi^2\alpha'^2 \mu_p}{12 g_s}  \int\dd^{p+1}\xi \sqrt{-(g+2\pi\alpha'\mathcal{F})} \biggl[ R_{\beta\delta}\left(\nabla_\alpha \mathcal{F}^{\alpha\beta} \nabla_\gamma \mathcal{F}^{\gamma\delta} - \nabla_\alpha \mathcal{F}_\gamma^{~\delta} \nabla^\gamma \mathcal{F}^{\alpha\beta}\right) \\
        &\qquad+ \frac{1}{2}R_{\beta\delta\gamma\epsilon}\nabla^\gamma \mathcal{F}^{\alpha\beta} \nabla^\epsilon  \mathcal{F}_\alpha^{~\delta} +\Omega^{a~\alpha}_{~\alpha}\nabla_\delta H_{\gamma~a}^{~\delta} \nabla_\beta \mathcal{F}^{\beta\gamma} \\ 
        & \qquad+ \frac{1}{4} R_\delta^{~\delta}\left(\nabla_\alpha \mathcal{F}^{\alpha\beta} \nabla_\gamma \mathcal{F}_\beta^{~\gamma} + \nabla_\beta \mathcal{F}_\alpha^{~\gamma} \nabla_\gamma \mathcal{F}^{\alpha\beta} \right)  \\
         &\qquad - \Omega^{a\beta\alpha} \left( \nabla_\beta \mathcal{F}_\alpha^{~\gamma} \nabla_\delta H_{\gamma~a}^{~\delta} + \nabla^\delta \mathcal{F}_\alpha^{~\gamma} \nabla_a H_{\beta\gamma\delta} - \frac{1}{2} \nabla^\delta \mathcal{F}_\alpha^{~\gamma}\nabla_\gamma H_{\beta\delta a} \right)\biggr]\,,
    \end{split}
\end{equation}
where $R_{\alpha\beta\gamma\delta}$ is the Riemann tensor and $R_{\alpha\beta}$ the Ricci tensor.
These terms also vanish since $F_2$ and $C_2$ are covariantly constant on the $S^2$ at the tip (property b)).
For the same reason the corrections in \cite{Garousi:2015mdg,BabaeiVelni:2016srs} vanish at the tip of the throat (except six terms of the form $\mathcal{F}_2\Omega^2\nabla H_3$ in \cite{BabaeiVelni:2016srs} that will be calculated in App.~\ref{sec:nonzeroflux}). 

Let us move on to corrections of the Chern-Simons (CS) action. The tree level Chern-Simons action reads
\begin{equation}
    S_{\text{CS,D}p}=\mu_p \int\limits_{\mathcal{M}_{p+1}}\left. \text{Tr}\,\left(\text{e}^{2\pi \alpha' \mathcal{F}_2}\right)\wedge \sqrt{\frac{\hat{A}(4\pi^2 \alpha' R_T)}{\hat{A}(4\pi^2 \alpha' R_N)}}\wedge \bigoplus\limits_q C_q\right|_{p+1}\,.
    \label{eq:csdptreelevel}
\end{equation}
This can be written in a more explicit form by expanding the exponential and the A-roof genus $\hat{A}(R_i)$ of the Riemann curvature 2-form of the normal or tangent bundle and then taking the right $C_q$ form such that the integrand matches a $p+1$ form. For the NS5-brane at the tip of the throat, the only non-vanishing components are $B_6 + \mathcal{F}_2\wedge C_4$ which are already included in the analysis of \cite{Kachru:2002gs}. Hence, there are no additional contributions from \eqref{eq:csdptreelevel} to the scalar potential.

Moreover, there are $\alpha'$ corrections to the CS action which are computed in \cite{Garousi:2010ki,Garousi:2010rn,Garousi:2011ut,Jalali:2015xca,Jalali:2016xtv}. 
When carefully taking into account properties a) to d), it turns out that all these terms vanish at the tip of the throat (except one term of the form $\epsilon \mathcal{F}_2 R \nabla \tilde{F}_5$ and two terms of the form $\epsilon \mathcal{F}_2  \nabla H_3 \nabla F_7$ of \cite{Jalali:2016xtv} which will be computed in App.~\ref{sec:nonzeroflux}).

\subsection{Non-zero flux corrections} \label{sec:nonzeroflux}

In \cite{Robbins:2014ara,Garousi:2014oya} $\alpha'^2$ corrections to the O-plane action are calculated. As explained in Sect.~\ref{sec:fluxcorretions} these are also present on D$p$-branes.
The corresponding action reads\footnote{As before, we assume that the couplings extend to non-geodesically embedded branes and abbreviate $H_3$ by $H$.}
\begin{equation}
    \begin{split}
    S_{\text{D$p$},\text{DBI}} \supset & \frac{\mu_p}{g_s}  \frac{\pi^2\alpha'^2}{48} \int \dd^{p+1}\xi \sqrt{-(g+2\pi\alpha'\mathcal{F}_2)} \biggl[  H^{\alpha\beta a} H_{\alpha~a}^{~\gamma}  (R_T)_{\beta\gamma} - \frac{3}{2} H^{\alpha\beta a} H_{\alpha\beta}^{~~~b} \, \overline{R}_{ab}\\
    &+\frac{1}{2} H^{abc}H_{ab}^{~~d}\,\overline{R}_{cd} 
     - H^{\alpha\beta a} H^{\gamma\delta}_{~~~a} (R_T)_{\alpha\beta\gamma\delta} + H^{\alpha\beta a} H_a^{~bc} (R_N)_{\alpha\beta bc}\\
     &- \frac{1}{4}  H^{\alpha\beta a} H_{\alpha\beta}^{~~~b} H_{a}^{~cd}H_{bcd} +\frac{1}{4}  H^{\alpha\beta a} H_{\alpha\beta}^{~~~b} H^{\gamma\delta}_{~~a} H_{\gamma\delta b} \\
    & +\frac{1}{8} H^{\alpha\beta a} H_\alpha^{~\gamma b} H_{\beta ~ b}^{~\delta} H_{\gamma\delta a} -\frac{1}{6} H^{\alpha\beta a} H_\alpha^{~\gamma b} H_{\beta\gamma}^{~~~c} H_{abc} +\frac{1}{24} H^{abc}H_a^{~de} H_{bd}^{~~f} H_{cef}
    \biggr]\,,
    \label{oplaneterms}
\end{split}
\end{equation}
where $(R_T)$, $(R_N)$ and $\overline{R}$ are defined as \cite{Bachas:1999um}
\begin{align}
    (R_T)_{\alpha \beta \gamma \delta} &= R_{\alpha \beta \gamma \delta} + g_{a b} (\Omega^a_{\alpha \gamma} \Omega^b_{\beta \delta} - \Omega^a_{\alpha \delta} \Omega^b_{\beta \gamma})\,,\\
    (R_N)_{\alpha\beta}^{\quad a b} &= -R^{a b}_{\,\,\,\,\alpha \beta} + g^{\gamma \delta} (\Omega^a_{\alpha \gamma} \Omega^b_{\beta \delta} - \Omega^b_{\alpha \gamma} \Omega^a_{\beta \delta})\,,\\
    \overline{R}_{a b} &= \hat{R}_{a b} + g^{\alpha \alpha'} g^{\beta \beta'} \Omega_{a \, \alpha \beta} \Omega_{b \, \alpha' \beta'}\,,
\end{align}
where $\Omega^\mu_{\alpha \beta}$ is the second fundamental form, and $\hat{R}_{a b} = R^\alpha_{\; a \alpha b}$.

In the case of the KS throat, we extend \eqref{oplaneterms} to NS5-branes as proposed in App.~\ref{sec:SdualAppendix} leading us schematically to the first two terms of \eqref{eq:dbins5} from which we can read off the correct $g_s$ scaling. Using the KS flux \eqref{eq:f3} in the spherical parametrization of the deformed conifold at the tip of the throat \cite{Nguyen:2019syc} (see also \cite{Hebecker:2022zme}), we find
\begin{equation}
    S_{\text{NS5},\text{DBI}} \supset -\frac{1}{(g_s M)^2}\left( c_3 + c_4 \cot^2\psi \right) \frac{\mu_5}{g_s^2}\int \dd^{6}\xi \sqrt{-(g+2\pi\alpha'g_s\mathcal{F}_2)}\,,
    \label{eq:dbif3}
\end{equation}
where $c_3= \frac{\pi^2}{48}\left(\frac{45}{4 I(0)^3}-\frac{12\,\, 6^{1/3}}{I(0)^2}\right)\approx2.44781$ and $c_4=\frac{\pi^2}{48}\frac{12\,\, 6^{1/3}}{I(0)^2}\approx8.69589$ with $I(0)\approx0.71805$. The coefficient $c_3$ is smaller than $c_4$ since the $F_3^4$ terms compete with the $F_3^2 R$ terms whereas $c_4$ is only due to $F_3^2\Omega^2$ couplings. 

In \cite{Garousi:2022rcv} couplings of the form $\Omega^4$ and $\Omega^4F_2^2$ are calculated where $\Omega$ is the second fundamental form. The couplings are (see equ.~(51) in \cite{Garousi:2022rcv})
\begin{equation}
\begin{split}
     S_{\text{D$p$},\text{DBI}} \supset - &\frac{\mu_p}{g_s}  \frac{\pi^2\alpha'^2}{24}\int \dd^{p+1}\xi \sqrt{-g} \biggl[9 \Omega^{a~\varepsilon}_{~\alpha} \Omega^{b~\eta}_{~\beta} \Omega_{b\gamma\varepsilon}\Omega_{a\delta\eta} (2\pi\alpha'F^{\alpha\beta}_2) (2\pi\alpha'F^{\gamma\delta}_2) \\
     & +2 \Omega_{a\gamma}^{~~\varepsilon} \Omega^{a\gamma\delta} \Omega^{b~\eta}_{~\delta} \Omega_{b\varepsilon\eta}  - 2 \Omega^{b}_{~\gamma\delta} \Omega^{a\gamma\delta} \Omega_{b\varepsilon\eta} \Omega_{a}^{~\varepsilon\eta}\\
     & + \frac{1}{4}\left( 2 \Omega_{a\gamma}^{~~\varepsilon} \Omega^{a\gamma\delta} \Omega^{b~\eta}_{~\delta} \Omega_{b\varepsilon\eta}  - 2 \Omega^{b}_{~\gamma\delta} \Omega^{a\gamma\delta} \Omega_{b\varepsilon\eta} \Omega_{a}^{~\varepsilon\eta} \right)(2\pi\alpha'F_{2\alpha\beta}) (2\pi\alpha'F_2^{\alpha\beta}) \\
     & + \underbrace{\dots}_\text{12 similar couplings with different contractions} + \mathcal{O}(\Omega^4 F_2^4) \biggr]\,.
\end{split}
\label{eq:omega4f2}
\end{equation}
The terms in the second line are precisely the pure $\alpha'^2$ curvature terms involving only the second fundamental form that were computed in \cite{Bachas:1999um}.
We observe that the index structure of the terms in the third line equals the index structure of the second line -- the field strength tensors are contracted among themselves. This means that we can identify the second and third line with the first two terms of the expansion of $\sqrt{-(g+2\pi\alpha'F_2)}\Omega^4$ which is already captured by the curvature correction of \cite{Bachas:1999um} calculated for the KS throat in \cite{Hebecker:2022zme}\footnote{Note that this is one reason why we assume that the couplings extend to $2\pi\alpha' F_2 \to 2\pi\alpha' \mathcal{F}_2$.}. 
Dropping the pure curvature terms, the novel coupling terms in \eqref{eq:omega4f2} read
\begin{equation}
\begin{split}
      S_{\text{D$p$},\text{DBI}} \supset - \frac{\mu_p}{g_s} & \frac{\pi^2\alpha'^2}{24}(2\pi)^2\int \dd^{p+1}\xi \sqrt{-(g+2\pi\alpha'\mathcal{F}_2)} \biggl[9 \Omega^{a~\varepsilon}_{~\alpha} \Omega^{b~\eta}_{~\beta} \Omega_{b\gamma\varepsilon}\Omega_{a\delta\eta}\mathcal{F}^{\alpha\beta}_2 \mathcal{F}^{\gamma\delta}_2 \\
     & + \underbrace{\dots}_\text{12 similar couplings with different contractions} \biggr]\,.
\end{split}
\label{eq:dbiomega4f2}
\end{equation}
Note that we extended the couplings by replacing $\sqrt{-g}\to\sqrt{-(g+2\pi\alpha'\mathcal{F}_2)}$ which means that we assume an infinite tower of (so far not calculated) couplings of the form $\Omega^4 \mathcal{F}_2^n$ rearranging in such a form to reproduce $
\sqrt{-(g+2\pi\alpha'\mathcal{F}_2)}$.
These terms can again be evaluated at the tip of the KS throat. Choosing the worldvolume $F_2$ flux as in KPV and the $g_s$ scaling as in \eqref{eq:dbins5}, they yield
\begin{equation}
    S_{\text{NS5},\text{DBI}} \supset 
    -\frac{\cot^4\psi \, c_5 }{(g_sM)^2 \sin^4\psi}\left(\frac{\pi p}{M}  -\left(\psi-\frac{\sin(2\psi)}{2}\right)\right)^2 \frac{\mu_5}{g_s^2}\int \dd^{6}\xi \sqrt{-(g+2\pi\alpha'g_s\mathcal{F}_2)}\,,
    \label{eq:dbiomega4f2ns5}
\end{equation}
where we wrote the $F_2$ flux number $p$ as $p= (p/M)(g_sM)/g_s$ and $c_5= \frac{\pi^2}{6} \frac{45\,\, 3^{1/3}}{4\,2^{2/3} I(0)^2}\approx 32.6096$. The parametrics of this result is understood easily: $\Omega^4\sim \cot^4\psi/R_{S^3}^4$, $F_2\sim p /R_{S^2}^2$ and $C_2\sim M(\psi-\sin(2\psi)/2)/R_{S^2}^2$ with $R_{S^2}^2= R_{S^3}^2\sin^2\psi\sim (g_sM)\sin^2\psi $.

Next, we evaluate the corrections to the DBI action calculated in \cite{BabaeiVelni:2016srs} at the tip of the throat. As explained above, the only non-vanishing corrections for the KS throat are given by
\begin{equation}
    \begin{split}
     S_{\text{D$p$},\text{DBI}} \supset - & \frac{\pi^2\alpha'^2\mu_p}{96 g_s }\int \dd^{p+1}\xi \sqrt{-(g+2\pi\alpha'\mathcal{F}_2)} \biggl[ 5 (2\pi\alpha'\mathcal{F}^{\alpha\beta}_2)\Omega^{a\gamma}_{~~\alpha} \Omega^{b~\delta}_{~\delta}\biggl( \nabla_a H_{\beta\gamma b} \\
     &- \nabla_b H_{\beta\gamma a} \biggr) 
    + (2\pi\alpha'\mathcal{F}^{\alpha\beta}_2)\Omega^{a\gamma}_{~~\alpha}\Omega^{b\delta}_{~~\gamma}\left( \nabla_a H_{\beta\delta b} - \nabla_b H_{\beta\delta a} \right) \\
    & + (2\pi\alpha'\mathcal{F}^{\alpha\beta}_2)\Omega^{a~\gamma}_{~\gamma}\Omega^{b~\delta}_{~\delta}\nabla_b H_{\alpha\beta a} - (2\pi\alpha'\mathcal{F}^{\alpha\beta}_2)\Omega^{b}_{\delta\gamma}\Omega^{a\delta\gamma}\nabla_b H_{\alpha\beta a}\biggr]\,.
\end{split}
\label{eq:dbif2omegah3}
\end{equation}
For the KS throat, the terms in the first two lines pairwise cancel against each other due to property c). The non-vanishing contribution then comes from the last line and yields for the NS5-brane
\begin{equation}
    S_{\text{NS5},\text{DBI}} \supset 
    \frac{\cot^3\psi \, c_6 }{(g_sM)^2 \sin^2\psi}\left(\frac{\pi p}{M}  -\left(\psi-\frac{\sin(2\psi)}{2}\right)\right)
    \frac{\mu_5}{g_s^2}\int \dd^{6}\xi \sqrt{-(g+2\pi\alpha'g_s\mathcal{F}_2)}\,,
    \label{eq:dbif2omegah3ns5}
\end{equation}
with $c_6  = \frac{6^{1/3}\pi^2 }{16 I(0)^2} \approx 2.17397$ where the parametrics can be understood from $g_s \nabla F_3\sim g_s M \cot\psi/R_{S^3}^4$.

Additionally to the non-vanishing contributions from the DBI action for the KS throat, we found one non-vanishing $\alpha'^2$ coupling from the Chern-Simons action that includes $\tilde{F}_5$ and two couplings that involve $H_7$. Both can be found in \cite{Jalali:2016xtv}. The $\tilde{F}_5$ coupling term on the NS5 is given by
\begin{equation}
    S_{\text{NS5},\text{CS}} \supset - \frac{\pi^2\alpha'^2 \mu_5}{24 g_s^2} \frac{1}{4!} \int\dd^{6}\xi \epsilon^{\alpha_0 \alpha_1\cdots \alpha_5} (2\pi\alpha'g_s\mathcal{F}_{2\,\alpha_0\alpha_1}) \overline{R}^{ab} \nabla_a (g_s\tilde{F}_5)_{b \alpha_2\alpha_3\alpha_4\alpha_5}\,,
    \label{eq:csf5}
\end{equation}
where $\epsilon^{\alpha_0 \alpha_1\dots \alpha_5}$ is the Levi-Civita symbol on the worldvolume of the NS5-brane.
This can be evaluated at the tip yielding
\begin{equation}
\begin{split}
     S_{\text{NS5},\text{CS}} \supset &-\left( - \frac{c_7 \left( \psi -\sin(2\psi)/2 \right)}{(g_sM)^2\sin^2\psi} + \frac{c_7 \pi (p/M) }{(g_sM)^2 \sin^2\psi}\right) \frac{\mu_5}{g_s^2}\int\dd^{6}\xi\sqrt{-g_6} \\
     & =-\frac{4\pi \mu_5 M}{g_s}\frac{c_7 }{(g_sM)^2}\left( \frac{\pi p}{M} - \left( \psi-\frac{\sin(2\psi)}{2}\right)\right) \int \dd^4 x \sqrt{-g_4}
\end{split}
\label{eq:csf5KS}
\end{equation}
where $I''(0)=- 2^{2/3}/3^{4/3}$ and $c_7= \frac{\pi^2\, 6^{2/3}\left( 2I(0) - 3 I''(0) \right)}{18 I(0)^{7/2}}\approx 14.6396$.

The $H_7$ coupling terms read (again already for the NS5 where $F_7\to H_7$)
\begin{equation}
\begin{split}
    S_{\text{NS5},\text{CS}} \supset  \frac{\pi^2\alpha'^2 \mu_5}{48g_s^2} \frac{1}{6!} & \int\dd^{6}\xi \epsilon^{\alpha_0 \alpha_1\cdots \alpha_5}\Biggl[ 
    (2\pi\alpha'g_s\mathcal{F}_2)^{\alpha\beta} \nabla^a(-g_s F_3)_{\alpha\beta}^{~~\,b} \nabla_b (g_s^2 H_7)_{a \alpha_0\cdots\alpha_5}\\
    & + (2\pi\alpha'g_s\mathcal{F}_2)^{\alpha\beta} \nabla^a(g_s F_3)_{\alpha\beta}^{~~~b} \nabla_a (g_s^2 H_7)_{b \alpha_0\cdots\alpha_5} \Biggr]\,,
    \end{split}
    \label{eq:csh7}
\end{equation}
which gives at the tip
\begin{equation}
      S_{\text{NS5},\text{CS}} \supset -
      \frac{4\pi \mu_5 M}{g_s}\frac{c_8 \cot\psi}{(g_sM)^2\sin\psi}\left(\frac{\pi p}{M} -\left( \psi -\frac{\sin(2\psi)}{2} \right) \right) \int \dd^4 x \sqrt{-g_4}\,,
      \label{eq:csh7KS}
\end{equation}
where $c_8 = 2\frac{\pi^2\, 6^{2/3}}{I(0)^{5/2}}\approx 18.6475$.

Taking all non-vanishing corrections together, the scalar potential following from 
\eqref{eq:dbif3}, \eqref{eq:dbiomega4f2ns5}, \eqref{eq:dbif2omegah3ns5}, \eqref{eq:csf5KS} and \eqref{eq:csh7KS} reads
\begin{equation}
\label{eq:vflux}
\begin{split}
    V_\text{flux} = & \frac{4 \pi \mu_5 M}{g_s} 
      \sqrt{b_0^4 \sin^4 (\psi) + \left(p \frac{\pi}{M} -\psi +\frac{1}{2}\sin(2\psi) \right)^2}
       \times\frac{1}{(g_sM)^2} \Biggl[ c_3\\
       & \qquad\qquad+ c_4 \cot^2\psi+\frac{ c_5 \cot^4\psi  }{ \sin^4\psi}\left(\frac{\pi p}{M}  -\left(\psi-\frac{\sin(2\psi)}{2}\right)\right)^2 \\
       & \qquad\qquad -\frac{c_6\,\cot^3\psi}{ \sin^2\psi}\left(\frac{\pi p}{M}  -\left(\psi-\frac{\sin(2\psi)}{2}\right)\right) \Biggr]\\
       & +\left[\frac{4\pi^2 p \mu_5}{g_s} -\frac{4 \pi \mu_5 M}{g_s} \left(\psi-\frac{\sin(2\psi)}{2}\right) \right] \left( \frac{c_7}{(g_sM)^2} + \frac{c_8 \cot\psi}{(g_sM)^2 \sin\psi}  \right)\,.
     \end{split}
\end{equation}

\section{Tadpole constraints in the Large Volume Scenario}
\label{sec:TadpoleAppendix}

In this appendix we first review the PTC \cite{Gao:2022fdi} and then give a reformulated version of it such that constraints from $\alpha'$ corrected KPV can naturally be implemented. In addition we briefly discuss a loop correction used to constrain the LVS in \cite{Junghans:2022exo,Junghans:2022kxg} and explain why we believe it to be less important than discussed in these papers.

\subsection{The Parametric Tadpole Constraint}\label{sec:sumptc}

To begin, we quickly review the basic set up of the LVS and the most important ingredients of the PTC. For more details consult \cite{Gao:2022fdi}. We work in type IIB string theory compactified to 4D on a Calabi-Yau orientifold. In the minimalist case, the Calabi-Yau has two Kahler moduli, a big cycle $\tau_b$ and a small cycle $\tau_s$. The volume is given by ${\cal V} = \tau_b^{3/2}-\kappa_s \tau_s^{3/2}$ in units of $l_s=2\pi\sqrt{\alpha'}$ where $\kappa_s=\sqrt{2}/(\sqrt{3}\kappa_{sss})$ and $\kappa_{sss}$ is the triple self intersection number of the small divisor.
The corresponding superpotential is given by
\begin{equation}
    W = W_0 + A_s \text{e}^{-a_s T_s}\,,
\end{equation}
where $\text{Re}(T_s)=\tau_s$ and $A_s$ a model-dependent prefactor. $W_0$ is induced by fluxes and $a_s$ determines whether the nonperturbative correction to $W$ is coming from ED3-branes ($a_s=2\pi$) or from gaugino condensation on D7-branes (for an $SO(8)$ gauge group, $a_s=\pi/3$). 
The Kahler potential reads \cite{Balasubramanian:2005zx,Becker:2002nn}
\begin{equation}
    K=-2\ln\left(\mathcal{V} + \frac{\xi}{2 g_s^{3/2}}\right) = -2 \ln \left( \tau_b^{3/2}-\kappa_s\tau_s^{3/2}-\frac{\chi\,\zeta(3)}{4(2\pi)^3g_s^{3/2}} \right)\,,
\end{equation}
where $g_s$ is the string coupling, $\chi$ the Euler number of the Calabi-Yau and $\zeta(3)\approx 1.2$. The correction proportional to $g_s^{-3/2}$ is due to \cite{Becker:2002nn}. 
At the minimum of the scalar potential 
\begin{equation}
    \label{vts}
    \mathcal{V}=\frac{3\kappa_s|W_0|\sqrt{\tau_s}}{4a_s |A_s|}\text{e}^{a_s\tau_s}\,\,\,,\qquad
    \qquad
    \tau_s =  \frac{\xi^{2/3}}{(2\kappa_s)^{2/3}g_s} +\mathcal{O}(1)\,.
\end{equation}

For deriving the PTC, higher F-term corrections \cite{Ciupke:2015msa,Junghans:2022exo} and a correction coming from the combination of higher curvature corrections and a varying warp factor \cite{Junghans:2022exo,Gao:2022fdi} are taken into account. Demanding the size of the corrections to be small compared to the LVS AdS minimum leads to the definition of the control parameters $c_{W_0}$ and $c_N$ which should be large for parametric control over the higher F-term or varying warp factor correction, respectively, \cite{Gao:2022fdi}:
\begin{equation}
    1 = c_{W_0} \frac{16 a_s }{3 (2\kappa_s)^{2/3}\xi^{1/3}}\, \frac{W_0^2}{\mathcal{V}^{2/3}}\,\,\,,\qquad\qquad 1= c_N \frac{10 \,a_s\,\xi^{2/3}}{(2\kappa_s)^{2/3}g_s} \frac{N}{\mathcal{V}^{2/3}} \,.
\end{equation}
Additionally to the PTC, there is a bound on $Q_3$ given by \cite{Denef:2004ze}
\begin{equation}
    -Q_3 \ge 4\pi\frac{g_s W_0^2}{2}\,,
    \label{denefdouglas}
\end{equation}
which can be used to determine $c_{W_0}^\ast(c_N)$ such that some minimal quality of control is ensured \cite{Gao:2022fdi}. This leads to $c_{W_0}^\ast(c_N)=15\pi\xi c_N/4$ such that the minimal tadpole required by the PTC reads\footnote{Note that we generalized the PTC for uplifting with $p$ anti-D3-branes.}
\begin{equation}
    -Q_3 > N = - \frac{21c_M}{16\pi} \mathcal{W}_{-1}(x)\,\,\,, \qquad x =- \frac{3^{11/35}\,a_0^{2/7}}{7\,\,2^{59/105}\, 5^{5/7}\, \pi^{1/35}} \, \frac{\kappa_s^{2/7}}{p^{2/7}\, a_s^{3/7}\, c_N^{5/7}\, c_M^{1/7} }
    \,,
    \label{ptc}
\end{equation}
where $\mathcal{W}_{-1}(x)$ is the $-1$ branch of the Lambert $\mathcal{W}$ function and we defined the control parameter $c_M=g_sM^2$.

\subsection{Loop Corrections} \label{sec:corrections}

In this section we will consider a loop correction to the LVS de Sitter uplift which was not taken into account when deriving the PTC but was considered in \cite{Junghans:2022exo,Junghans:2022kxg} and we will discuss the relative importance of this loop correction.

The Kahler potential for the Kahler moduli will generically be loop corrected, see e.g.~\cite{vonGersdorff:2005bf,Berg:2005ja,Berg:2005yu,Berg:2007wt,Cicoli:2007xp,Cicoli:2008va}. A detailed study of such loop corrections and their relative importance was recently given in \cite{Gao:2022uop}.

We will focus only on the most dangerous loop corrections which are loop corrections to the blowup cycle as they are in the LVS minimum only suppressed by $g_s^2$ (instead of the volume $g_s^{3/2}\mathcal{V}^{-1/3}$) compared to the leading order terms in the scalar potential \cite{Gao:2022uop}. 

From this it directly follows that loop corrections in the LVS will be small as soon as $g_s^2$ is sufficiently small. This can be made more precise by formulating this condition in terms of a control parameter. A reasonable way to define the control parameter is to measure the size of the loop correction to the blowup cycle compared to its leading order value \eqref{vts}. Using the results of \cite{Junghans:2022exo} this can for instance be specified for a `KK-type'\footnote{We refer the reader to \cite{Gao:2022uop} for a classification of the different types of loop corrections.} loop correction to the Kahler potential of the form \cite{Berg:2007wt}
\begin{equation}
    \delta K =C_s^{\text{KK}} \frac{g_s\sqrt{\tau_s}}{\mathcal{V}}\,,
    \label{loopcskk}
\end{equation}
where $C_s^\text{KK}$ is some unknown prefactor. This leads to a correction\footnote{Note that also winding type loop correction to the blow-up cycle will contribute at the same order in $g_s$.} to the value of $\tau_s$ in the LVS minimum in \eqref{vts} of the form $C_s^\text{KK} g_s/(3\kappa_s)$. 
The control parameter is hence defined as 
\begin{equation}
    1 = c_\text{loop}\, \frac{2^{2/3}g_s^2}{3\kappa_s^{1/3}\xi^{2/3}}\,,
    \label{cloop}
\end{equation}
and we demand $c_\text{loop} \gg 1$ for control. This is a much weaker constraint than the related one of \cite{Junghans:2022kxg}. They demand $\lambda_6=a_sg_s/(3\kappa_s) \ll 1$.
This requirement follows from demanding that the size of the loop correction is smaller than one, such that the correction to the vev of the volume modulus $\Delta\mathcal{V}=\mathcal{V}\exp(a_s g_s/(3\kappa_s))$ is small. 
In our opinion, a correction to the volume modulus is only dangerous if the new value of the volume is no longer at exponentially large values. This only happens when the correction is comparable to the leading order term which leads to \eqref{cloop}. If the volume gets corrected by $\Delta\mathcal{V}=\mathcal{V}\exp(\mathcal{O}(1))$, we assume that one can tune the parameters $W_0$, $N$, and $g_s$ of the model such that a different dS minimum for the new vevs can be found. We therefore propose to use \eqref{cloop} instead of $\lambda_6$. 

Furthermore, as recently discussed in \cite{Gao:2022uop}, there are settings where loop corrections are assumed to be absent. For example, the loop correction \eqref{loopcskk} is absent if no D7-brane wraps the blowup cycle $\tau_s$. Then, a `KK-type' loop corrections can not be induced since the $M_{10}^2g_sR_8^2$ operator on the D7 and an Einstein-Hilbert term on an intersection 2-cycle $\sim \sqrt{\tau_s}$ is absent \cite{Gao:2022uop}. Additionally to \eqref{loopcskk} there are `Winding-type' loop corrections \cite{Berg:2007wt} or more generally genuine loop corrections \cite{Gao:2022uop} to the blowup cycle. Such corrections appear more generally and are (probably) only absent if there is $\mathcal{N}=2$ SUSY locally at the blowup cycle. 
Hence, it should be possible to find models not featuring loop corrections to the small blowup cycle. 

As discussed in \cite{Gao:2022uop}, one may attempt to estimate the numerical prefactor in the expansion parameter of loop corrections and hence their $1/2\pi$ suppression. Such a suppression can play a major role since \eqref{cloop} would be weakened significantly such that parametric control over loop corrections could be obtained much earlier than expected. 
For some toric geometries the expansion parameter is calculated explicitly in \cite{Berg:2005ja} and a $1/(2\pi)^4$ suppression is found. For generic Calabi-Yaus naive dimensional analysis \cite{Chacko:1999hg} in a 4d approach reveals a suppression by $1/16\pi^2$ \cite{Gao:2022uop} without evaluating the sum over KK modes which can lead to a further suppression (see e.g.~Sect.~II in \cite{Cheng:2002iz})\footnote{Note that there exists a counterexample where the loop correction is only suppressed in $g_s$. This is the loop correction to the 10d $R^4$ term calculated in \cite{Antoniadis:1997eg}. We thank Daniel Junghans for bringing this to our attention.}.

As these loop corrections are $g_s^2$ suppressed and we expect them to be suppressed by a numerical prefactor generically no larger than $1/16\pi^2$, we shall assume they can usually be neglected in our analysis. In cases where $g_s\leq 1$ for some choice of $M$ and one does not want to rely on a small numerical prefactor of loop corrections, it is still possible to increase $M$ and keep $g_sM$ constant to decrease $g_s$. This is of course at the expense of increasing the D3-tadpole.

\subsection{The bound from $g_s M$} \label{sec:gsmptc}

The constraints coming from the $\alpha'$ corrected KPV potential can naturally be thought of as a bound on $g_s M$ since $g_sM$ is the main control parameter of the potential. 

We hence aim to rewrite \eqref{ptc} such that it takes $c_e \equiv g_sM$ instead of $c_M=g_sM^2$ as an input parameter. As usual, we treat $c_e$ as a control parameter where large $c_e$ increases the control over $\alpha'$ corrections. This can be done using\footnote{Note that $c_e$ depends on $c_N$ through $N$ logarithmically. With increasing $c_N$ also $c_e$ decreases.}
\begin{equation}
    c_e = g_sM=\sqrt{g_s c_M} = \left(\frac{9 a_s}{16\pi}\right)^{1/2} \left(\frac{\xi}{2\kappa_s}\right)^{1/3} \frac{c_M}{\sqrt{N}}\,, 
    \label{ceofcm}
\end{equation}
where we used
\begin{equation}
    g_s = \frac{9 a_s}{16\pi} \left(\frac{\xi}{2\kappa_s}\right)^{2/3} \frac{c_M}{N}\quad \Longleftrightarrow\quad \frac{g_s}{p} = \frac{9 a_s}{16\pi} \left(\frac{\xi}{2\kappa_s}\right)^{2/3} \frac{c_e }{  N}\frac{M}{p}\,,
    \label{gs}
\end{equation}
following from a relation $\tau_s(N)$ derived in \cite{Gao:2022fdi} from comparing the volume in \eqref{vts} with the volume expressed in terms of the uplift potential.

The version of the PTC which takes $c_e$ instead of $c_M$ as input is then derived by using \eqref{ceofcm} in \eqref{ptc}. We end up with
\begin{equation}
     \hspace*{-.3cm}-Q_3 > N_{\{c_N,c_e\}} = \frac{2^{8/3}}{\pi} \frac{\kappa_s^{2/3} \, c_e^2 }{\xi^{2/3} \, a_s} \left( \mathcal{W}_{-1}(y)\right)^2 ,\,\,  y= - \frac{3^{2/5}\,a_0^{1/4}}{8\,\,2^{23/40}\, 5^{5/8}\, \pi^{1/40}} \, \frac{\kappa_s^{1/6}\, \xi^{1/12}}{p^{1/4}\, a_s^{1/4}\, c_N^{5/8}\, c_e^{1/4} }
    \,.
    \label{ptc2}
\end{equation}

Let us now compare the two versions of the PTC \eqref{ptc} and \eqref{ptc2}. In \eqref{ptc} the only parameter that enters non-logarithmically is $c_M$ whereas in \eqref{ptc2} $c_e$ and the topological quantities enter polynomially. 
If one considers \eqref{ptc} with a given $c_M$ and topological quantities, the value of $c_e$ is then determined by \eqref{ceofcm}, which we call $c_e^\ast$. This value of $c_e^\ast$ can be in contradiction with the constraints of Fig.~\ref{fig:poverm} on $c_e$ obtained from the $\alpha'$ corrected KPV potential. The contradiction arises if $c_e^\ast$ is smaller than the  value obtained from Fig.~\ref{fig:poverm} since then the KPV potential does not have a metastable minimum any more. In the KPV context it is hence more useful to consider the PTC \eqref{ptc2} instead of \eqref{ptc} in order to obtain bounds on the minimal tadpole.

\section{$\alpha'$ corrections on NS5-branes} \label{sec:SdualAppendix}

In this Appendix we explain in detail how to obtain $\alpha'$ corrections on NS5-branes from $\alpha'$ corrections on D-branes. The main logic goes as follows. First, we make use of the fact that the D3-brane is self-dual under S-duality to all orders in $\alpha'$. Due to this property, the subleading $g_s$ corrections on the D3-brane at order $\alpha'^2$ can be inferred. Second, we assume that, according to the Myers effect \cite{Myers:1999ps}, the on-shell action of a fluxed D5-brane is equivalent to a non-abelian stack of D3-branes in the limit when the $S^2$ on which the D5-brane is wrapped shrinks to zero size. From this we obtain the subleading $g_s$ corrections at order $\alpha'^2$ on the D5-brane. Finally, we S-dualize the D5-brane action including the subleading $g_s$ corrections to obtain the $\alpha'^2$ corrections on NS5-branes. The subleading $g_s$ corrections are crucial since S-duality relates terms at different order in $g_s$.

The above argument relies on the following assumptions:
\begin{itemize}
    \item[1)] We assume that the D3-brane is self-dual under S-duality to all order in $\alpha'$. This is only proven to leading order in $\alpha'$ in \cite{Tseytlin:1996it,Green:1996qg,Kimura:1999jb} but a derivation including $\alpha'$ corrections is to our knowledge missing.
    \item[2)] The $\alpha'$ and $g_s$ corrected action of a fluxed D5-brane should provide the same physics as the $\alpha'$ and $g_s$ corrected action of a nonabelian stack of D3-branes due to the Myers effect. In particular we expect the on-shell action in both perspectives to match when they describe the same physics. To show this assumption one would have to study the action of a nonabelian brane stack at higher order both in $\alpha'$ and commutators. We leave studying the physics of a nonabelian brane stack at higher orders as an interesting direction for future work.
    \item[3)]
    We should note that point 2) above rests on a crucial further assumption\footnote{We thank an anonymous referee for pointing this out to us.}. Namely that when shrinking the $S^2$ on which the D5-brane is wrapped, one should recover the D3 action order by order in $\alpha'$ from the D5 action. This is not guaranteed for the following reason: 
    In the Myers effect, the D3-brane and D5-brane effective actions are controlled in opposite regimes.
    On the one hand the nonabelian effective action for a stack of $p$ D3-branes can only be used when the distance between the D3-branes is substringy\footnote{Strictly, as discussed for the D0-D2 Myers effect in \cite{Myers:1999ps}, it is only necessary that the branes nearest each other are a substringy distance removed for control in the nonabelian perspective. Applying this to the D3-D5 system, the radius over which a nonabelian stack of $p$ D3-branes can be spread in a controlled manner is then bounded as $R<\sqrt{p} \,l_s$. For sufficiently large $p$ there is then an intermediate radius where both the nonabelian D3 and the fivebrane perspective can be trusted. However, this does not resolve the issue, fundamental to us, that control in the fivebrane is lost if one shrinks the $S^2$ that the fivebrane wraps to zero size.}. Otherwise the lightest open string states stretching between different branes would have mass $m > 1/l_s$. Hence, in order to obtain the D3-brane action from the D5 action, one should reduce the D5-brane worldvolume on a substringy $S^2$. On the other hand, the D5-brane effective action clearly holds only when $\text{vol}(S^2) > l^2_s$. It could then be that the $\alpha'^2$ terms in the D3-brane effective action get contributions from an infinite set of higher-derivative terms supported by the D5-brane, and not just from the corresponding D5-brane $\alpha'^2$ terms as we will assume in the following. Note that at leading order in $\alpha'$ starting from the D5 action and shrinking the $S^2$ to zero size, one does exactly obtain the D3 action at leading order in $\alpha'$, making it not completely implausible that this matching between the actions will continue to hold order by order in $\alpha'$.

    \item[4)] All $\alpha'^2$ corrections to the DBI and CS action have subleading corrections in $g_s$ as given by the same Eisenstein series. The form of the Eisenstein series is convincingly stated for the case of curvature corrections in \cite{Bachas:1999um,Green:2000ke,Basu:2008gt}. The argument is extended to some flux corrections in \cite{Garousi:2011fc}. Therefore it seems reasonable to assume that the Eisenstein series multiplies each $\alpha'^2$ correction even though the subleading corrections in $g_s$ have not been proven so far. 
\end{itemize}

Let us now explain this logic in more detail by focusing on the curvature corrections for explicitness. According to assumption 3) this logic then applies to all other $\alpha'$ corrections.

To begin, we consider the self-dual DBI action of the D3-bane in Einstein frame. It is given by \cite{Bachas:1999um,Green:2000ke,Basu:2008gt,Garousi:2011fc}
\begin{equation}
    S_{\text{D3,EF}} \supset \frac{\mu_3 \pi^2\alpha'^2}{24} \int_{\mathcal{M}_{4}} \dd^{4}x\sqrt{-g} E_1(\tau,\overline{\tau}) R^2\,,
    \label{eq:d3sinv}
\end{equation}
where we suppressed all different contractions of the $R^2$ terms and neglected numerical prefactors. Crucially, however, the non-holomorphic Eisenstein series $E_1(\tau,\overline{\tau})$ appears in front of all contractions in the same way. The Eisenstein series enjoys an expansion in $\tau_2=\text{Im} \tau$ (see e.g.~\cite{Basu:2008gt,Garousi:2011fc}):
\begin{equation}
    E_1(\tau,\overline{\tau}) = \frac{\tau_2}{2} - \frac{\pi}{4\zeta(2)}\ln(\tau_2) +\frac{\pi}{2\zeta(2)} \sqrt{\tau_2} \sum_{m\neq0,n\neq0} \left| \frac{m}{n} \right|^{1/2} K_{1/2}(2\pi|mn|\tau_2)\text{e}^{2\pi i mn \tau_1}\,,
    \label{eq:eisenstein}
\end{equation}
where $\zeta(z)$ is the zeta function. The first term corresponds to the well known tree level term, the $\ln\tau_2$ term will be canceled by anomaly cancellation as explained in \cite{Bachas:1999um}, and the last term denotes non-perturbative instanton corrections where $K_{1/2}(x)$ is a Bessel function.

The next step is to propose that the subleading corrections in $g_s$ on a fluxed D5-brane are also given by the function $E_1(\tau,\overline{\tau})$. The reason is that due to the Myers effect, the action of a fluxed D5-brane should correspond to the action of a nonabelian stack of coincident D3-branes when shrinking the $S_2$ wrapped by the D5-brane to zero size.
This can be shown easily for the tree level action (see e.g.~\cite{Hebecker:2022zme} for an explicit calculation) but we assume that this also holds at higher order in $\alpha'$ \footnote{The higher order nonabelian D3-brane stack computation which is required for comparison becomes significantly more complex than the tree-level analysis. We hope to discuss the higher-order physics of a nonabelian stack in future work.}
Then, the $\alpha'^2$ curvature corrections to all orders in $g_s$ on the D5-brane in Einstein frame read
\begin{equation}
    S_{\text{D5,EF}} \supset \frac{\mu_5\pi^2 \alpha'^2}{24}g_s^{1/2} \int_{\mathcal{M}_{4}\times S_2}\dd^{6}x\sqrt{-g+\mathcal{F}_2}\,E_1(\tau,\overline{\tau})\, R^2\,,
    \label{eq:curvd5}
\end{equation}
where we again neglected numerical prefactors and the different contractions of the $R^2$ terms.
This action is, as expected, not invariant under S-duality due to the factor of $g_s^{1/2}$. 
The curvature corrections on the NS5-brane are then obtained by S-dualizing the corrections \eqref{eq:curvd5} on the D5-brane. 
According to this logic, the curvature corrections on the NS5-brane read in string frame
\begin{align}
    S_{\text{NS5,SF}} \supset&\, \frac{\mu_5 \pi^2\alpha'^2}{24g_s^2} \int_{\mathcal{M}_{4}\times S_2}\dd^{6}x\sqrt{-g+\mathcal{F}_2}\,g_s\,E_1(\tau,\overline{\tau})\, R^2\\ 
    = & \,\frac{\mu_5\pi^2 \alpha'^2}{24g_s^2} \int_{\mathcal{M}_{4}\times S_2}\dd^{6}x\sqrt{-g+\mathcal{F}_2}\left(\frac{1}{2}+\cdots\right) R^2\,,
\end{align}
where we expanded $E_1(\tau,\overline{\tau})$ to leading order in $g_s$ according to \eqref{eq:eisenstein}.
The leading order term in $g_s$ therefore scales as the tension term of the NS5-brane. 

From this we conclude that all $\alpha'^2$ corrections appear at tree level in $g_s$ and can be obtained from corrections on the D5-brane by replacing the fields by their S-dual counterpart.

\subsection{Naively S-dualizing the D5-brane action}

One may wonder why we went through the process of first deriving the D5 action to all orders in $g_s$ and then S-dualizing. Why not simply S-dualize the tree level in $g_s$ but $\alpha'$-corrected D5 action and use the NS5-brane action obtained in this way? The reason is that as stated previously, S-dualizing exchanges different orders in $g_s$. In this appendix we will examine in some more detail what such a naive analysis misses compared to a complete analysis using all order, perturbative and non-perturbative, in $g_s$.

For instance, S-dualizing the tree level in $g_s$ but $\alpha'^2$ terms of the D5 action leads to $g_s^2$ suppressed terms compared to the tree level NS5-brane action. S-dualizing corrections at tree level in $g_s$ therefore effectively produces two loop open string effects on the NS5-brane. This can for instance be shown for the $\alpha'^2$ corrections given in \eqref{eq:dbidp} where S-dualizing for $p=5$ leads to
\begin{equation}
\begin{split}
    S_{\text{DBI,NS}5} \supset \mu_5\alpha'^2 &\int_{\mathcal{M}_6}\dd^{6}x \sqrt{-(g+2\pi\alpha' g_s\mathcal{F}_2)} 
    \Biggl[ (-g_sF_3)^4 + (-g_sF_3)^2 R \\
    &+ \Omega^4(2\pi\alpha'g_s\mathcal{F}_2)^2 + (2\pi\alpha'g_s\mathcal{F}_2)\Omega^2 \nabla(-g_sF_3) \Biggr]\,.
    \end{split}
\end{equation}
The overall prefactor is $g_s$ independent (each $F_3$ and $\mathcal{F}_2$ always comes with a factor $g_s$ as explained in Sect.~\ref{sec:fluxcorretions}). Hence the terms are $g_s^2$ suppressed compared to the tension term of the NS5-brane which has a $1/g_s^2$ dependence.

In order to obtain the correct $g_s$ scaling when shrinking the $S_2$ wrapped by the NS5-brane to zero size we then have to  propose that there should be also a term at tree level in $g_s$ with the same numerical prefactor. According to the expansion of the Eisenstein series \eqref{eq:eisenstein} this leads to a conundrum since in the expansion there is no term scaling like a two loop effect. Resolving this issue is the subject of the next section.

\subsection{Concerning $\alpha'$ corrections and S-duality}


In this appendix we wish to make a few clarifying statements concerning the interpretation of the interplay between S-duality and $\alpha'$ corrections to the spacetime and brane actions in IIB string theory. Our discussion of the reinterpretation of instanton effects as partially resumming to provide effective perturbative $g_s$ loop corrections has to our knowledge not appeared in the literature although some aficionados are likely already implicitly aware of these results.

For simplicity of exposition, we will not consider the full $SL(2,\mathbb{Z})$ of S-duality but restrict ourselves to $C_0=0$ and consider the strong-weak duality $g_s \rightarrow g_s'=1/g_s$.

Let us consider first the D3-brane, which is assumed to be invariant under S-duality. The D3-brane action contains an $\alpha'^2$ curvature correction which reads in string frame at leading order in $g_s$
\begin{equation}
    S_{\text{D3,curv}} = \frac{\mu_3\pi^2 \alpha'^2 }{48}\int_{\mathcal{M}_{4}} \dd^{4}x\sqrt{-g} \, \frac{1}{g_s} R^2\,.
\end{equation}
One may S-dualize this action to obtain
\begin{equation}
    S_{\text{D3,curv}} = \frac{\mu_3\pi^2 \alpha'^2 }{48}\int_{\mathcal{M}_{4}} \dd^{4}x\sqrt{-g} \,g_s' \, R^2\,.
\end{equation}
Clearly this new action is not the same as the original action, it is $g_s^2$ suppressed. It is then tempting to conclude that in order for the action to be self-dual under S-duality, the full term taking into account all orders of $g_s$ should be 
\begin{equation}
\label{eqcurvnaivepert}
    S_{\text{D3,curv}} = \frac{\mu_3\pi^2 \alpha'^2 }{48}\int_{\mathcal{M}_{4}} \dd^{4}x\sqrt{-g}\,\left(g_s^{-1} + c + g_s^{1}\right) R^2\,.
\end{equation}
However, this conclusion is incorrect as it does not account for non-perturbative in $g_s$ instanton corrections and, falsely, assumes an existing two loop term. In fact, the full $\alpha'^2$ curvature corrections to all orders in $g_s$, both perturbative and non-perturbative is known to be \cite{Bachas:1999um,Green:2000ke,Basu:2008gt,Garousi:2011fc}
    \begin{equation}
    \label{eqcurvweisen}
    S_{\text{D3,curv}} = \frac{\mu_3\pi^2\alpha'^2}{24} \int_{\mathcal{M}_{4}} \dd^{4}x\sqrt{-g} \, E_1(\tau,\overline{\tau}) \,R^2\,,
\end{equation}
where we neglected the contractions of the $R^2$ terms and $E_1(\tau,\overline{\tau})$ is a non-holomorphic Eisenstein series.
The Eisenstein series can be expanded as \cite{Basu:2008gt,Garousi:2011fc}
\begin{equation}
    E_1(\tau,\overline{\tau}) = \frac{\tau_2}{2} - \frac{\pi}{4\zeta(2)}\ln(\tau_2) +\frac{\pi}{2\zeta(2)} \sqrt{\tau_2} \sum_{m\neq0,n\neq0} \left| \frac{m}{n} \right|^{1/2} K_{1/2}(2\pi|mn|\tau_2)\text{e}^{2\pi i mn \tau_1}\,,
    \label{eq:eisenstein1}
\end{equation}
which shows that there is no two loop term but an infinite series of non-perturbative in $g_s$ instanton corrections.

We therefore conclude that in the limit of large $g_s$, the infinite series of non-perturbative corrections effectively scales at leading order as a two loop term. Explicitly plotting $E_1$ indeed confirms this behaviour. For large $g_s$, the instanton series will then yield the leading term in $g_s'$ under S-duality. In the language of naively S-dualizing this corresponds to the statement that a two loop term in $g_s$ S-dualizes to a tree level term in $g_s'$. 

If for $C_0=0$ one defines $f(g_s)\equiv 2 E_1(g_s)-g_s-g_s^{-1}$ then one could write \eqref{eqcurvweisen} as 
\begin{equation}
    S_{\text{D3,curv}} = \frac{\mu_3\pi^2 \alpha'^2 }{48}\int_{\mathcal{M}_{4}} \dd^{4}x\sqrt{-g}\,\left(g_s^{-1} + f(g_s) + g_s^{1}\right) R^2\,,
\end{equation}
where $f(g_s)$ is a function which falls to zero both in the limit $g_s\rightarrow 0$ and $g_s \rightarrow +\infty$. This differs from \eqref{eqcurvnaivepert} obtained from the perturbative analysis in that the middle term is not a constant but a function $f(g_s)$ set by the non-perturbative effects. This function matters at finite $g_s$ but disappears in the asymptotic limits.

The same logic applies to the $\alpha'^3 R^4$ corrections of type IIB string theory where naively S-dualizing the tree level in $g_s$ correction produces a non-existing $g_s^3$ suppressed term on the S-dual side. Taking into account the full $\alpha'^3$ curvature corrections to all order in $g_s$ in form of an Eisenstein series of weight $3/2$ \cite{Green:1997tv}, one finds again that at large $g_s$ the non-perturbative corrections sum up to provide a behaviour which at leading order in the large $g_s$ limit scales as a $g_s^3$ suppressed term would.

To conclude, naive S-dualization of the leading order term in $g_s$ provides the exact behaviour in the large $g_s$ limit even when that behaviour actually arises through the summation over an infinite number of non-perturbative corrections.

\bibliographystyle{JHEP}
\bibliography{refs}

\end{document}